\documentclass[aps,prx,twocolumn,superscriptaddress]{revtex4-2}

\usepackage[T1]{fontenc}
\usepackage[separate-uncertainty=true]{siunitx}
\usepackage[mathlines]{lineno}
\usepackage{gensymb}
\usepackage{graphicx}
\graphicspath{{Figures/}}		
\usepackage{amssymb,amsmath,bbm,bm}
\usepackage{float}
\usepackage{braket}
\usepackage[colorlinks=true, linkcolor=blue,  citecolor=blue,
urlcolor=blue]{hyperref}

\usepackage{xcolor}

\newcommand{\figpanel}[1]{(#1)}

\newcommand{\cavpuls}{\omega_{c}}
\newcommand{\drivepuls}{\omega_{d}}
\newcommand{\qubitpuls}{\omega_{q}}
\newcommand{\detuning}{\epsilon_\delta}
\newcommand{\tunnel}{t_{\rm c}}
\newcommand{\gtrans}{g_{\perp}}
\newcommand{\glong}{g_{\parallel}}
\newcommand{\glongavg}{\bar{g}_{\parallel}}
\newcommand{\glongmod}{\tilde{g}_{\parallel}}
\newcommand{\gbare}{g_0}
\newcommand{\detqd}{\Delta_{qd}}
\newcommand{\detcd}{\Delta_{cd}}

\newcommand{\phiLong}{\phi_\parallel}
\newcommand{\phiRadiation}{\phi_m}
\newcommand{\phiTunnel}{\phi_t}
\newcommand{\readoutPhase}{\varphi}
\newcommand{\phiInterf}{\delta \phi}

\newcommand{\driveAmp}{A_{in}}
\newcommand{\driveFast}{A_{g}}
\newcommand{\driveCav}{A_{d}}

\newcommand{\ann}[1]{\hat{#1}}
\newcommand{\cre}[1]{\hat{#1}^\dagger}
\newcommand{\numop}[1]{\hat{#1}^\dagger\hat{#1}}

\newcommand{\avg}[1]{\langle #1 \rangle}
\NewDocumentCommand{\ketbra}{m O{}}{
  \left| \IfBlankTF{#2}{#1}{#2} \right\rangle \!\! \left\langle #1 \right|
}

\begin{document}

\title[Parametric drive of a double quantum dot in a cavity]{Parametric drive of a double quantum dot in a cavity}

\author{L.~Jarjat}
\affiliation{Laboratoire de Physique de l’\'{E}cole normale sup\'{e}rieure, ENS, Universit\'{e} PSL, CNRS, Sorbonne Universit\'{e}, Universit\'{e} Paris Cit\'{e}, Paris, France}

\author{B.~Hue}
\affiliation{Laboratoire de Physique de l’\'{E}cole normale sup\'{e}rieure, ENS, Universit\'{e} PSL, CNRS, Sorbonne Universit\'{e}, Universit\'{e} Paris Cit\'{e}, Paris, France}
\affiliation{C12 Quantum Electronics, Paris, France}

\author{T.~Philippe-Kagan}
\affiliation{Laboratoire de Physique de l’\'{E}cole normale sup\'{e}rieure, ENS, Universit\'{e} PSL, CNRS, Sorbonne Universit\'{e}, Universit\'{e} Paris Cit\'{e}, Paris, France}

\author{B.~Neukelmance}
\affiliation{Laboratoire de Physique de l’\'{E}cole normale sup\'{e}rieure, ENS, Universit\'{e} PSL, CNRS, Sorbonne Universit\'{e}, Universit\'{e} Paris Cit\'{e}, Paris, France}
\affiliation{C12 Quantum Electronics, Paris, France}

\author{J.~Craquelin}
\affiliation{Laboratoire de Physique de l’\'{E}cole normale sup\'{e}rieure, ENS, Universit\'{e} PSL, CNRS, Sorbonne Universit\'{e}, Universit\'{e} Paris Cit\'{e}, Paris, France}

\author{A.~Théry}
\affiliation{Laboratoire de Physique de l’\'{E}cole normale sup\'{e}rieure, ENS, Universit\'{e} PSL, CNRS, Sorbonne Universit\'{e}, Universit\'{e} Paris Cit\'{e}, Paris, France}

\author{C.~Fruy}
\affiliation{Laboratoire de Physique de l’\'{E}cole normale sup\'{e}rieure, ENS, Universit\'{e} PSL, CNRS, Sorbonne Universit\'{e}, Universit\'{e} Paris Cit\'{e}, Paris, France}

\author{G.~Abulizi}
\affiliation{C12 Quantum Electronics, Paris, France}

\author{J.~Becdelievre}
\affiliation{C12 Quantum Electronics, Paris, France}

\author{M.M.~Desjardins}
\affiliation{C12 Quantum Electronics, Paris, France}

\author{T.~Kontos}
\affiliation{Laboratoire de Physique de l’\'{E}cole normale sup\'{e}rieure, ENS, Universit\'{e} PSL, CNRS, Sorbonne Universit\'{e}, Universit\'{e} Paris Cit\'{e}, Paris, France}
\affiliation{Laboratoire de Physique et d'\'{E}tude des Mat\'{e}riaux, ESPCI Paris, Universit\'{e} PSL, CNRS, Sorbonne Universit\'{e}, Paris, France}

\author{M.R.~Delbecq}
\affiliation{Laboratoire de Physique de l’\'{E}cole normale sup\'{e}rieure, ENS, Universit\'{e} PSL, CNRS, Sorbonne Universit\'{e}, Universit\'{e} Paris Cit\'{e}, Paris, France}
\affiliation{Laboratoire de Physique et d'\'{E}tude des Mat\'{e}riaux, ESPCI Paris, Universit\'{e} PSL, CNRS, Sorbonne Universit\'{e}, Paris, France}
\affiliation{Institut universitaire de France (IUF)}

\begin{abstract}

We demonstrate the parametric modulation of a double quantum dot charge dipole coupled to a cavity, at the cavity frequency, achieving an amplified readout signal compared to conventional dispersive protocols. Our findings show that the observed cavity field displacement originates from dipole radiation within the cavity, rather than from a longitudinal coupling mechanism, yet exhibits the same signatures while relying on a transverse coupling. By carefully tuning the phase and amplitude of the intra-cavity field, we achieve a $\pi$-phase shift between two dipole states, resulting in a substantial enhancement of the signal-to-noise ratio. In addition to its applications in quantum dot based qubits in cQED architectures, this protocol could serve as a new promising tool for probing exotic electronic states in mesoscopic circuits embedded in cavities.
\end{abstract}

\maketitle


\section{Introduction}

Cavity quantum electrodynamics (CQED) is a versatile platform for manipulating various qubits, including Rydberg atoms~\cite{Haroche2006}, superconducting circuits~\cite{Wallraff2004,blais2020,blais2021}, and semiconductor quantum dots~\cite{Xiang2013,Burkard2020,Clerk2020}. Central to CQED and circuit QED (cQED) is the dipole interaction with the cavity field~\cite{Haroche2006}, enabling quantum information processing and providing a powerful probe of condensed matter physics~\cite{Cottet2017}. Regardless of application, improving the signal-to-noise ratio (SNR) is essential. This work focuses on enhancing readout performance in hybrid cQED architectures, with broader implications for exploring exotic electronic phenomena.

\begin{figure}[hbt]
\centering
\includegraphics[width=0.8\linewidth, angle=0]{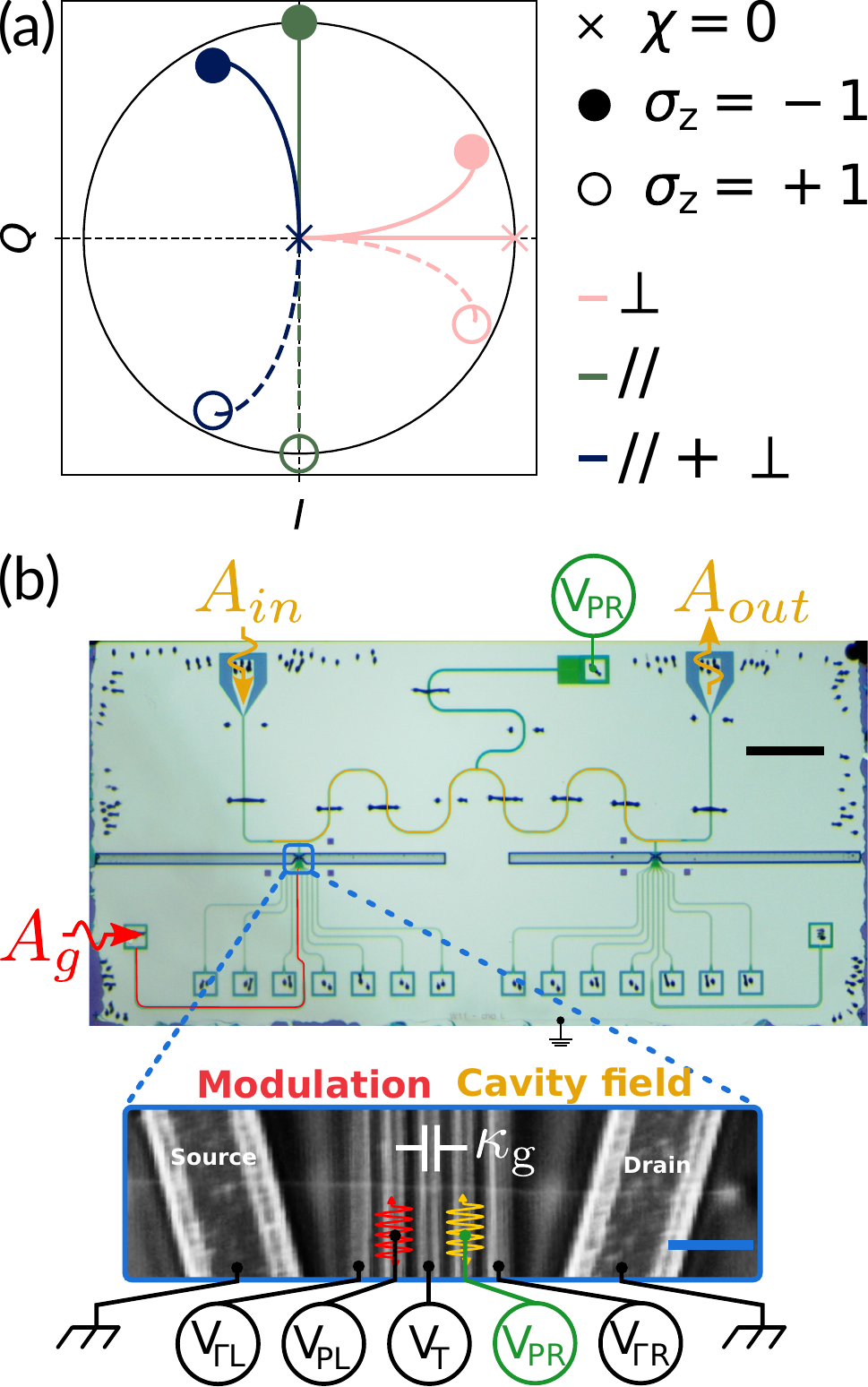}
\caption{Principle of the experiment.
\figpanel{a} Cavity field state-dependent trajectories for $\detcd=0$ and cases $\sigma_z=\pm 1$ and $\chi=0$. The pure transverse coupling is in pink, the pure parametric longitudinal coupling is in green and the combination of both is in dark blue. The steady states are shown as dots at the end of the trajectories and the reference states for $\chi=0$ are shown as crosses. \figpanel{b} Top: optical image of the device showing the Nb resonator (yellow) and the parametric drive line (red). The scale bar is \SI{1}{\milli\metre}. Bottom: SEM image of a typical DQD device with a transferred CNT. The scale bar is \SI{500}{\nano\metre}.
}
\label{fig1:principle}
\end{figure}

For a conventional transverse dipole-cavity coupling, the interaction Hamiltonian reads $\gtrans \ann{\sigma}_x (\ann{a} + \cre{a})$, with $\ann{a}$ the cavity mode annihilation operator, $\ann{\sigma}_i$ Pauli matrices in the qubit space, and $\gtrans$ the transverse dipole-photon coupling. The cavity field drives the dipole, which shifts the cavity frequency by a state-dependent term $\chi \avg{\sigma_z}$, with $\chi=\gtrans^2 / (\detqd - i \Gamma_2)$, where $\detqd=\qubitpuls - \drivepuls$ is the qubit-drive detuning, and $\Gamma_2$ the dipole decoherence rate. Under a drive $\hbar |A_{in}| [e^{-i (\drivepuls t +\phi_{in})}\ann{a}+h.c.]$, the cavity field is displaced along the pink curved trajectories in Fig.~\ref{fig1:principle}\figpanel{a}, with a steady-state value

\begin{equation}\label{eq:field_disp}
\bar{a} = \frac{A_{in} e^{-i \phi_{in}}}{-\detcd + i \frac{\kappa}{2}-\chi \avg{\sigma_z}},
\end{equation}

where $\detcd=\cavpuls - \drivepuls$ is the cavity-drive detuning and $\kappa$ the cavity loss rate. A straightforward improvement of the SNR typically involves increasing the cavity photon number $n_{ph}=\avg{\numop{a}}$, but it also enhances qubit dephasing via transverse coupling~\cite{Clerk2010}.

To overcome these limitations, parametric longitudinal coupling has been proposed~\cite{Kerman2013,Didier2015,Lambert2017,bosco2022,chessari2024}, introducing an interaction $\glong \hat{\sigma}_z (\ann{a} + \cre{a})$, where $\glong = \glongavg + \glongmod \cos(\cavpuls t + \phiLong)$. This adds a term $\glongmod \avg{\hat{\sigma}_z} e^{-i \phiLong}$ to the numerator of eq.~\eqref{eq:field_disp}. Without transverse coupling, this results in linear state-dependent trajectories in phase space, whereas a residual transverse coupling produces curved trajectories (respectively green and blue trajectories in Fig.~\ref{fig1:principle}\figpanel{a}). Even with residual coupling, the field separation is larger and grows faster than with pure dispersive readout, improving readout SNR and speed, assuming equal photon numbers in all schemes (Fig.~\ref{fig1:principle}\figpanel{a}). Beyond qubit readout, parametric longitudinal coupling could facilitate the detection and manipulation of exotic electronic states~\cite{Contamin2021} and enable direct electronic charge measurements instead of compressibility~\cite{Desjardins2017}.

In hybrid cQED with double quantum dots (DQDs), the cavity-dipole interaction naturally includes transverse and longitudinal terms given by $\gtrans=\gbare \sin \theta$ and $\glong = \gbare \cos \theta$, where $\gbare$ is the bare DQD dipole-photon coupling, and $\theta=\mathrm{arctan}(2 \tunnel / \detuning)$ is the DQD pairing angle, with $\tunnel$ the tunnel coupling and $\detuning$ the energy detuning between the dots~\cite{childress2004}. Note that $\gtrans$ ($\glong$) is maximal (zero) at $\detuning=0$ and vanishing (maximal) for $\detuning \gg \tunnel$. The coupling $\gbare$ arises from a charge density coupling to the electric part of the cavity field~\cite{Cottet2015}. Consequently, it seems natural to implement the longitudinal coupling readout by modulating $\detuning$ via a local electrostatic gate drive $V_g \cos(\omega t + \phiRadiation)$ as in refs.~\cite{corrigan2023,champain2024}. However, this necessarily produces a parasitic cavity drive due to capacitive coupling between the device gates, as shown in Fig.~\ref{fig1:principle}\figpanel{b}. Thus, eq.~\eqref{eq:field_disp} is modified to 

\begin{equation}\label{eq:field_full}
	\bar{a} = \frac{\driveCav + A_{\parallel} + A_{mod}}{-\Delta_{cd} + i \frac{\kappa}{2}-\chi \avg{\sigma_z}},
\end{equation}

\sloppy comprising three drive terms in the numerator (see Supplemental Material). The first term corresponds to the interferences between the drives from the input and the parasitic ports of the cavity, $\driveCav=|A_d|e^{-i \phi_{d}}=(|\driveAmp| + |\driveFast|e^{-i \delta \phi})e^{-i \phi_{in}}=|\driveAmp|e^{-i \phi_{in}}(1 + r e^{-i \delta \phi})$. The second term is the longitudinal coupling term $A_\parallel=-\glongmod \avg{\sigma_z}e^{-i\phiLong}$. Finally there is a third term coming from the direct modulation of the dipole, \(A_{mod}=\tilde{\chi} (\beta_L e^{-i \phiRadiation} \allowbreak - \mathrm{cotan} \theta \beta_t e^{-i \phiTunnel})\avg{\sigma_z}\). The two terms in $A_{mod}$ correspond respectively to the modulation of $\detuning$ and $\tunnel$, with $\beta_L$ ($\beta_t$) the electrostatic lever arm to $\detuning$ ($\tunnel$) from the modulation gate and $\tilde{\chi}=q V_g /(2 \hbar) \times \chi/\gbare$ with $q$ the electric charge defining the dipole. All terms have a phase factor that cannot be ignored as we are now effectively dealing with a multiple interference situation.

The parametric readout protocol a priori requires to cancel out the direct cavity drive term $A_d$, as done in ref.~\cite{champain2024} and similarly in an effective implementation with superconducting qubits~\cite{Touzard2019}. Here, we go further by fully exploiting the amplitude and phase tuning of $A_d$ to elucidate the underlying mechanism at play. In all the measurements of this work, we keep $\avg{\sigma_z}=-1$ (thermal ground state), and instead adjust $\detuning$, which controls both $\theta$ and $\chi$ through $\qubitpuls=\sqrt{\detuning^2+4 \tunnel^2}$. This strategy, combined with precise control of $A_d$, allows us to clearly identify the dominant parametric contribution.

\section{Results}

We implement this parametric scheme in a device similar to the one shown in Fig.~\ref{fig1:principle}\figpanel{b}, made of a microwave Nb coplanar waveguide resonator coupled to a suspended carbon nanotube (CNT) based DQD~\cite{Cubaynes2020,neukelmance2024}. The DQD is electrostatically defined with five gate electrodes. The voltages applied to the coupling gates to the left and right contacts are $V_{\rm \Gamma L}=V_{\rm \Gamma R}=\SI{-1}{\volt}$ and the tunnel gate is set to $V_{\rm T}=\SI{-1}{\volt}$ while both contacts are grounded throughout. All experiments are performed at a temperature of \SI{20}{\milli\kelvin}. The detuning $\detuning$ is controlled by tuning the voltages on the left and right plunger gate voltages. The cavity is coupled to the DQD via the right plunger gate, while the parametric drive is applied via the left plunger gate. The cavity features an input port $\kappa_{in}$ and an output port $\kappa_{out}$ for standard readout measurements. Mutual capacitance between the DQD plunger gates creates an additional input port, $\kappa_g$ (see Fig.~\ref{fig1:principle}(b)), through which the cavity field is directly driven along with the parametric DQD modulation we investigate.

\begin{figure}[hbt]
\centering
\includegraphics[width=0.98\linewidth, angle=0]{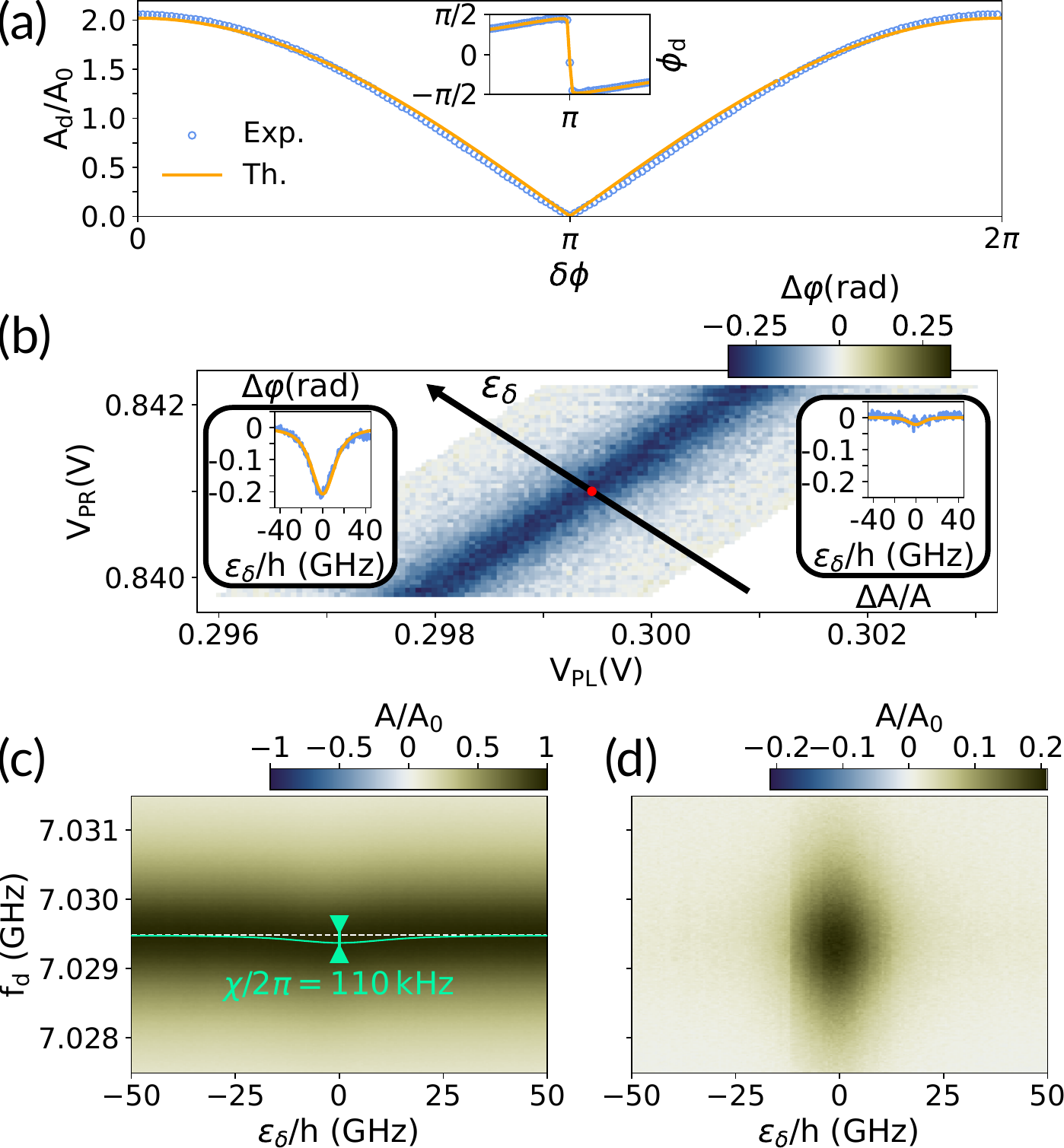}
\caption{Dipole radiation in the cavity. 
\figpanel{a} Normalized output cavity field amplitude $A_d/A_0$, for $\chi=0$ ($\detuning \gg \tunnel$) and $r\approx 1$, as a function of the phase difference $\phiInterf$ between the input ($A_{in})$ and parasitic ($A_g$) drives. Inset: phase $\phi_d$ of $A_d$ around $\phiInterf=\pi$. \figpanel{b} Phase variation $\Delta \readoutPhase$ of the cavity field probed at $f_c$ across an interdot transition, as a function of the left and right plunger gate voltages. Insets: cuts along the detuning axis (black arrow) of $\Delta \readoutPhase$ (left) and $\Delta A / A_{\rm off}$ (right). Normalized output field amplitude as a function of detuning $\detuning$ and drive frequency $f_d$ in the standard dispersive readout regime \figpanel{c} and with parametric modulation of $\detuning$ at $f_d$ for $r=0.995$ and $\phiInterf=\pi$ \figpanel{d}.
}
\label{fig2:dipole_radiation}
\end{figure}

We drive the cavity through $\kappa_{in}$ or $\kappa_g$, giving in both cases the same transmission in amplitude and phase (see Supplemental Material). The cavity frequency is $f_c=\cavpuls/2\pi=\SI{7.0295}{\giga\hertz}$ with linewidth $\kappa/2\pi=\SI{1.34}{\mega\hertz}$. We set $r=|\driveFast|/|\driveAmp|\approx 1$ and find $\kappa_g \approx \kappa_{in} / 1665$. We then tune the relative phase $\phiInterf$ between $\driveAmp$ and $\driveFast$ to reveal the intra-cavity interferences. We show in Fig.~\ref{fig2:dipole_radiation}\figpanel{a} the normalized amplitude $\driveCav/A_0$ (with $A_0$ the transmitted amplitude at resonance when driving through $\kappa_{in}$ with the dipole turned off at $\chi=0$ for $\detuning \gg \tunnel$) and phase $\phi_d$ at $r=0.995$. We observe intra-cavity field cancellation which sets the value $\phiInterf=\pi$.

Next, we tune the DQD electric dipole. In Fig.~\ref{fig2:dipole_radiation}\figpanel{b}, we show the phase variation $\Delta \readoutPhase=\readoutPhase - \readoutPhase_{\rm off}$, with $\readoutPhase_{\rm off}$ the reference phase at $\detuning \gg \tunnel$ (or $\chi = 0$), of the transmitted microwave field at the cavity frequency $f_c$, driven exclusively through $\kappa_{in}$, as a function of both plunger gate voltages, revealing an interdot dipole transition. The parameters of this dipole transition are determined through temperature dependence measurements and two-tone spectroscopy (see Supplemental Material), yielding $\tunnel \approx \SI{8.8}{\giga\hertz}$, giving $\qubitpuls / (2\pi) \approx \SI{17.6}{\giga\hertz}$, and $\gbare \approx \SI{31}{\mega\hertz}$. By defining the detuning as $\detuning = q(\beta_L V_{\rm PL} - \beta_R V_{\rm PR})$, we extract the electrostatic lever arms $\beta_L = -0.091$ and $\beta_R = 0.113$, assuming $q = -|e|$ with $e$ as the elementary charge. Next, we measure the normalized transmitted microwave amplitude $A / A_0$ along the detuning axis, indicated by the black arrow in Fig.~\ref{fig2:dipole_radiation}\figpanel{b}, and as a function of the drive frequency $f_d$, as shown in Fig.~\ref{fig2:dipole_radiation}\figpanel{c}. As expected, the cavity transmission shows barely noticeable distortion, even near $\detuning = 0$, where the dipole is maximal, since we have $\chi / 2\pi \approx \SI{110}{\kilo\hertz} \ll \kappa / 2 \pi$.

The situation drastically changes when simultaneously driving through $\kappa_{in}$ and $\kappa_{g}$ with $r\approx 1$ and $\phiInterf=\pi$, canceling the intra-cavity field, as shown in Fig.\ref{fig2:dipole_radiation}\figpanel{d}. Under these conditions, a finite field amplitude is observed only when the drive frequency $f_d$ matches the cavity frequency $f_c$ and when the dipole moment is simultaneously maximal near $\detuning = 0$. No field is present at large $\detuning$. This measurement provides insights into the dominant term in eq.~\eqref{eq:field_full} based on the parity of the signal with respect to $\detuning$, which is clearly even. The term $A_\parallel \propto \detuning / \sqrt{\detuning^2 + 4 \tunnel^2}$ is odd in $\detuning$, as is the $\beta_t$ term in $A_{mod}$, which is proportional to $2\detuning \tunnel / (\detuning^2 + 4 \tunnel^2)$. In contrast, the $\beta_L$ term in $A_{mod}$, proportional to $4\tunnel^2 / (\detuning^2 + 4 \tunnel^2)$, is even in $\detuning$. Therefore, the intra-cavity field mainly originates from this $\beta_L$ term, corresponding to the dipole radiation within the cavity.

\begin{figure}[hbt]
\centering
\includegraphics[width=1.0\linewidth, angle=0]{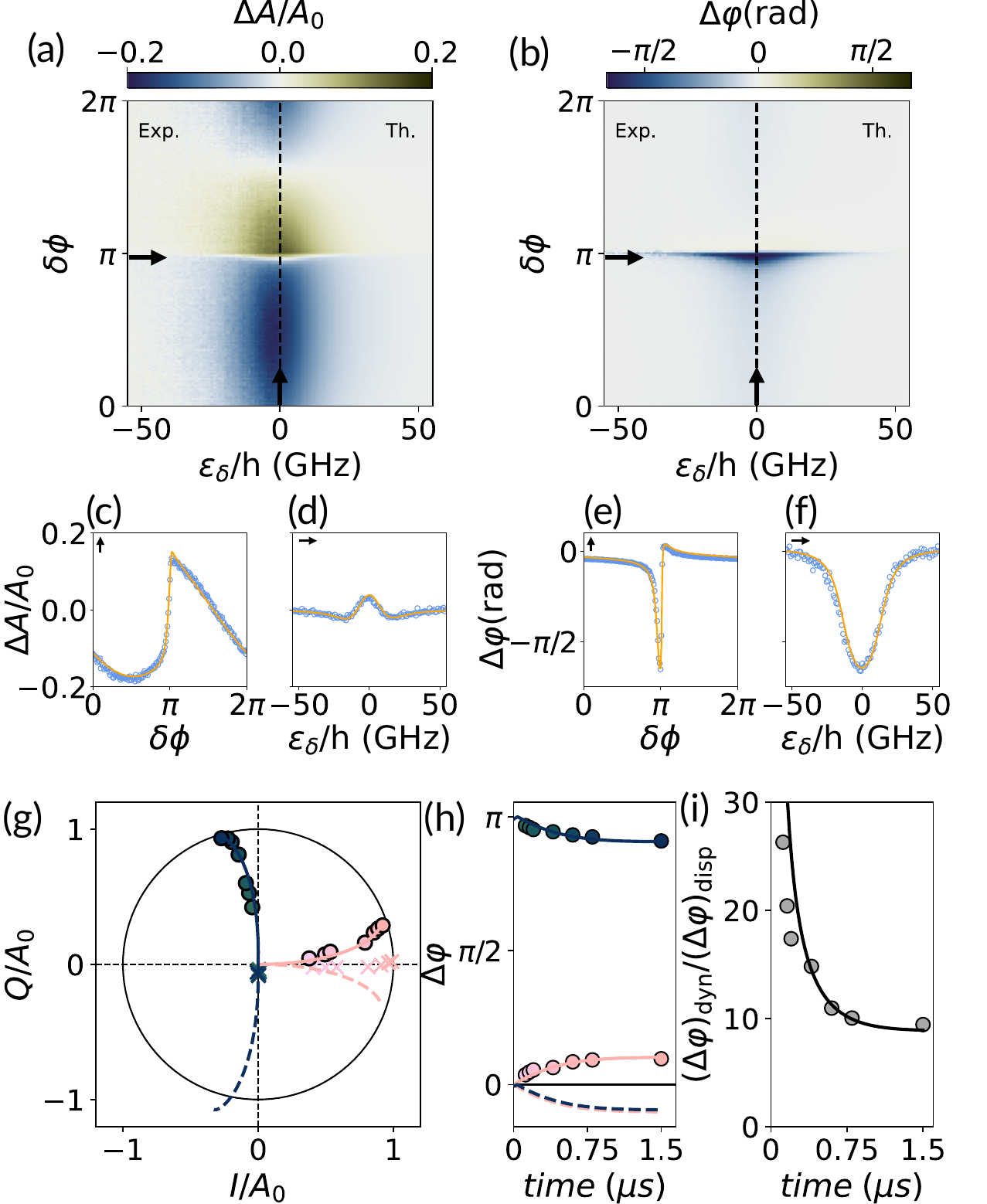}
\caption{Dynamical readout of the interdot transition.
\figpanel{a} Relative field amplitude variation $\Delta A / A_0$ of the interdot transition taken along the detuning axis (black arrow in Fig.~\ref{fig2:dipole_radiation}\figpanel{b}) and varying the drives interference phase $\phiInterf$ at constant $r\approx 1$. Experimental data (left side) and theory (right side) are displayed on the same plot. \figpanel{b} Same as \figpanel{a} but for the quantity $\Delta \readoutPhase$. Respective cuts along $\phiInterf$ at zero detuning are shown in \figpanel{c,e} and cuts along $\detuning$ at $\phiInterf\approx0.99 \pi$ are shown in \figpanel{d,f}. Data are shown by blue circles and simulations by orange lines. (g) IQ plane trajectories as a function of readout time for the dispersive (light pink) and dynamic (dark blue) readout. Corresponding reference states are shown as crosses. The solid and dashed lines are theory for the $\avg{\sigma_z}=-1$ and $+1$ respectively. (h) Corresponding phase contrast $\Delta \varphi$ and (i) their ratio for the two readout schemes.
}
\label{fig3:dynamical_readout}%
\end{figure}

To further investigate the intra-cavity field displacement, we fix the drive frequency to $f_d = f_c$ and instead vary $\phiInterf$ in order to fully determine the field drive terms with their associated phases. The results are presented in Fig.~\ref{fig3:dynamical_readout}, showing experimental data for $\detuning < 0$ and simulations for $\detuning > 0$ in panels \figpanel{a} and \figpanel{b}, respectively, for $\Delta A / A_0$ and $\Delta \readoutPhase$, with $\Delta A = A - A_{\rm off}$. Corresponding cuts along $\phiInterf$ and $\detuning$ are shown in panels \figpanel{c-f}, demonstrating excellent quantitative agreement between experiment and simulation. Similar agreement holds for cases $r\approx1.2$ and $r\approx0.6$, where we see that the sign of $\Delta \readoutPhase$ can change from negative to positive value (see Supplemental Material). Here, with $r\approx 1$, $\Delta \readoutPhase$ already reaches a maximum contrast value close to \SI{2}{\radian}, approximately 10 times larger than in the conventional dispersive readout shown in Fig.~\ref{fig2:dipole_radiation}\figpanel{b}. This corresponds to a tenfold SNR improvement at equal field amplitude.

The odd part in $\detuning$ of the data (see Supplemental Material) confirms that $A_\parallel$ is negligible, as a finite longitudinal term would yield a clear antisymmetric signal at large $\detuning$. Thus, we can already conclude that the parametric modulation of $\detuning$ does not induce a longitudinal coupling modulation. For simulations, we set $A_\parallel=0$ and fix $\phi_m=\phi_t$ based on the odd part of the data. We fit the data using only $\phi_m$, $\beta_t$, and $V_{g}$ as free parameters, obtaining $\beta_t\approx 0.03 \beta_L$ and $\phi_m \approx 0.7 \pi$. Given that the DQD modulation and the parasitic drive originate from the same gate, one might expect $\phi_m$ closer to zero. However, measurements on a different dipole transition in the same device yield $\phi_m\approx0.5 \pi$, suggesting that $\phi_m$ depends on the DQD parameters rather than mere propagation phase. A dynamic inductive or capacitive response of the DQD can also be ruled out, as it would exhibit a $\detuning$ dependence~\cite{frey2012a,Bruhat2016} that we do not observe, leaving the exact origin open for further investigation. Finally, to verify that the modulation remains adiabatic, we evaluate the non-adiabaticity parameter $\eta = \varepsilon \drivepuls / \qubitpuls^2$~\cite{shevchenko2010}, where $\varepsilon = \beta_L q V_g / 2\hbar$ is the drive amplitude. For our parameters, we find $\eta \approx 0.04$, indicating that the system operates well within the adiabatic regime. This excludes non-linear effects such as Landau-Zener transitions and supports the assumption that $\langle \sigma_z \rangle$ remains fixed during the parametric drive.

To complete our investigation of the dynamical readout scheme, we now analyze the trajectory of the cavity field in the IQ plane as a function of readout time, from \SI{0.12}{\micro\second} to \SI{1.5}{\micro\second} (Fig.~\ref{fig3:dynamical_readout}\figpanel{g}). We compare the dynamical readout scheme (dark blue dots) to the conventional dispersive readout (light pink dots) for a second interdot transition with parameters $\tunnel \approx \SI{6.3}{\giga\hertz}$ and $\gbare \approx \SI{29}{\mega\hertz}$ (see Supplemental Material). Reference trajectories for $\chi=0$ (obtained at $\epsilon_\delta \gg t_c$) are marked by crosses and reveal distinctly different behaviors: in the dynamical scheme, the reference state remains close to the origin, while in the dispersive case it moves linearly towards maximum amplitude. The measured trajectories agree well with predictions (solid lines for our case $\avg{\sigma_z}=-1$ and dashed lines for the case $\avg{\sigma_z}=1$, not studied here) of Fig.~\ref{fig1:principle}\figpanel{a}.
Importantly, the parametric drive trajectory follows the trajectory predicted for the longitudinal readout scheme with residual transverse coupling (Fig.~\ref{fig1:principle}\figpanel{a}) while we observed the absence of longitudinal coupling signal in Fig.~\ref{fig3:dynamical_readout}\figpanel{a,b}. This confirms the crucial role of our experimental approach, fixing $\langle\sigma_z\rangle=-1$ and varying $\chi$ via $\detuning$ in unambiguously identifying dipole radiation as the mechanism behind cavity-field displacement when modulating $\detuning$, since IQ trajectories alone cannot distinguish between longitudinal and dipole radiation mechanisms. Finally, Fig.~\ref{fig3:dynamical_readout}\figpanel{h} compares the phase contrasts $\Delta\varphi$ as a function of readout time for both dispersive and dynamical schemes, while Fig.~\ref{fig3:dynamical_readout}\figpanel{i} displays their ratio. These data highlight a clear enhancement of readout signal with the dynamical approach, from roughly 25 at short times to about 10 at long readout times for this dipole, demonstrating a substantial readout advantage.

\begin{figure}[hbt]
\centering
\includegraphics[width=0.9\linewidth, angle=0]{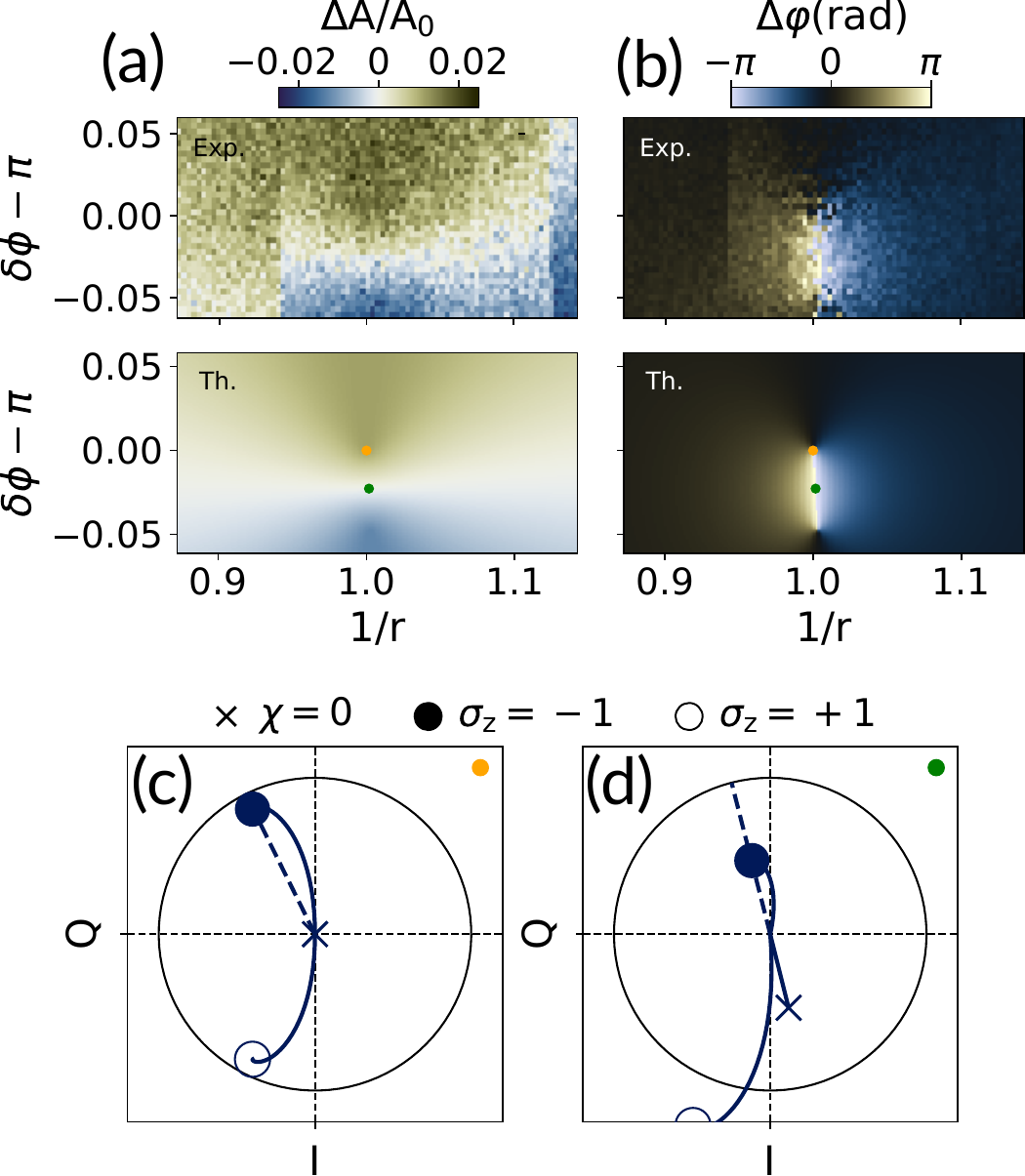}
\caption{Optimization of the readout phase contrast and enhancement of the SNR. 
Normalized amplitude variation $\Delta A / A_0$ \figpanel{a} and phase variation $\Delta \readoutPhase$ \figpanel{b} of the output cavity field as a function of the amplitude imbalance $1/r$ between the input and parasitic drives and relative phases $\phiInterf$, close to $\pi$. The reference field output is taken at $\chi=0$ (large $\detuning$). Top plots show the experimental data and bottom plots show the theoretical simulations. \figpanel{c,d} Cavity field phase space trajectories and steady-state positions at the orange and green spots shown in \figpanel{a,b}. Here we measure the amplitude and phase variation between the cases $\avg{\sigma_z}=-1$ and $\chi=0$.
}
\label{fig4:SNR}
\end{figure}

We now turn to the optimization of the SNR in the steady state by tuning the amplitude and phase of the direct cavity drive. Figure~\ref{fig4:SNR}\figpanel{a,b} shows the normalized transmitted field amplitude and phase variations, $\Delta A / A_0$ and $\Delta \readoutPhase$, as functions of the drive amplitude ratio $r$ and relative phase $\phiInterf$, with again quantitative agreement with theory simulations. The reference field corresponds to the dipole being turned off at $\detuning \gg \tunnel$, where $\chi = 0$. These measurements were done on a third dipole transition with parameters $\tunnel \approx \SI{20.8}{\giga\hertz}$ and $\gbare \approx \SI{22}{\mega\hertz}$ (see Supplemental Material) which results in a maximum phase contrast in the conventional dispersive regime of about \SI{0.05}{\radian}.

When the direct cavity drive field is canceled ($A_d = 0$), the situation depicted in Fig.\ref{fig4:SNR}\figpanel{c} is achieved, indicated by an orange dot in the bottom panels of Fig.~\ref{fig4:SNR}\figpanel{a,b}. The phase at the ``off'' point ($\chi=0$) is undefined because the cavity is empty so that $\Delta \readoutPhase$ gives no information, while $\Delta A$ reaches its maximum. We achieve the maximum phase contrast of $\pi$ at the green dot in Fig.~\ref{fig4:SNR}\figpanel{d}, where the two measurement points (``on'' and ``off'') are symmetrically centered around the origin of the IQ plane. This condition is satisfied when $A_d = A_{mod} / (2 - 2 \chi / \kappa)$, leading to an SNR improvement by a factor of about 60 for this dipole. In future experiments, it should be straightforward to implement a configuration where the separation between the two conditionally displaced fields for $\sigma_z = \pm 1$ is maximized. This can be achieved by tuning $r$ and $\phiInterf$ such that $A_d = A_{mod} \times i 2 \chi / \kappa $.

\section{Discussion}

In this work, we demonstrated that parametric modulation of the detuning in a DQD coupled to a cavity does not modulate the bare electron-photon coupling $\gbare$, as required for ``true'' longitudinal readout~\cite{Didier2015}. However, the cavity field is still conditionally displaced based on the dipole state encoded in $\sigma_z$, exhibiting identical trajectories to the longitudinal scheme. Thus, this approach significantly enhances the readout SNR and speed, making it directly applicable to hybrid cQED qubits based on DQDs~\cite{Mi2018,Samkharadze2018,Borjans2020,dijkema2024,neukelmance2024}. This enhancement is important for faster measurement times, advancing towards single-shot readout. Importantly, genuine longitudinal readout should operate at large $\detuning$, where charge noise is significant, whereas our dynamical readout occurs at zero detuning, where sensitivity to charge noise is minimal.

The dipole radiation mechanism identified in this work arises from transverse coupling. Although it significantly improves the SNR, it does not offer the full benefits of true longitudinal coupling, such as photon-induced dephasing elimination and quantum non-demolition measurements~\cite{Didier2015,Billangeon2015}. Furthermore, only a genuine parametric longitudinal scheme can realize two-qubit gates without virtual photons, thus potentially reducing decoherence and operation times~\cite{Kerman2013}. Achieving modulation of the longitudinal coupling therefore remains a desirable goal. This could be realized by  modulating the impedance of the resonator transmission line.

Finally, the dipole radiation scheme demonstrated here could effectively be used to investigate exotic electronic states. For example, with Majorana zero modes (MZM), the longitudinal coupling term between a MZM pair and the cavity field is proportional to their energy overlap $\varepsilon$~\cite{Cottet2015}. A local parametric drive of $\varepsilon$ might thus modulate the longitudinal coupling, allowing detection and manipulation of MZMs~\cite{Contamin2021}. Additionally, this readout technique could measure the absolute charge of a dipole: conventional dispersive readout would first calibrate $\chi$ and $\beta_L$, while subsequent parametric modulation could directly determine the dipole charge $q$ from $\tilde{\chi}$. This could notably be used to detect fractional charges in topological quantum dot chains~\cite{Park2016}.
\newline\newline
This work was supported by the French National Research Agency (ANR) through the grant MITIQ (TK), the ANR JCJC STOIC (ANR-22-CE30-0009) (MRD), by the ANR through the France 2030 programme through the PEPR MIRACLEQ (ANR-23-PETQ-0003) (MRD) and by the BPI through grant QUARBONE (TK).
\newline\newline
\emph{Data availability}---The data that support the findings of this manuscript are openly available~\cite{DataRepo}.

%


\clearpage
\onecolumngrid
\begin{center}
\textbf{\Large Supplementary Materials for ``Parametric drive of a double quantum dot in a cavity''}
\end{center}

\setcounter{equation}{0}
\setcounter{section}{0}
\setcounter{figure}{0}
\setcounter{table}{0}
\setcounter{page}{1}

\renewcommand{\theequation}{S\arabic{equation}}
\renewcommand{\thesection}{S\arabic{section}}

\renewcommand{\figurename}{Supplementary Figure}
\renewcommand{\tablename}{Supplementary Table}

\title{\Large Supplementary Materials for ``Microsecond-lived quantum states in a carbon-based circuit driven by cavity photons''}

\maketitle

\section{DQD-cavity coupled system with parametric modulation of the detuning}

The Hamiltonian of the DQD-cavity coupled system in the $\{\ket{L}, \ket{R} \}$ basis of left and right dots with single particle occupation is

\begin{equation}\label{eq:full_H_no_mod}
	\hat{H} = \hat{H}_{\rm cav} + \hat{H}_{\rm dqd} + \hat{H}_{\rm int} + \hat{H}_{\rm drive}
\end{equation}

with

\begin{align*}
	\hat{H}_{\rm cav} & = \hbar \cavpuls \left(\numop{a} + \frac{1}{2}\right) \\
	\hat{H}_{\rm dqd} & = \epsilon_L \ketbra{L} + \epsilon_R \ketbra{R} + \tunnel \ketbra{R}[L] + \tunnel \ketbra{L}[R] \\
	\hat{H}_{\rm int} & = \hbar \left(g_L \ketbra{L} + g_R \ketbra{R}\right) \left(\ann{a} + \cre{a} \right) \\
	\hat{H}_{\rm drive} & = \hbar |\driveAmp| \left[e^{i (\drivepuls t + \phi_{in})} \ann{a} + e^{-i (\drivepuls t + \phi_{in})} \cre{a} \right],
\end{align*}

where $\epsilon_{L/R}$ are the energies in each state $\ket{L/R}$. The electron-photon couplings to each dot are given by~\cite{Cottet2015}

\begin{equation}
    g_{L/R} = -|e| \int d\vec{r}^3 V_\perp(\vec{r}) \Psi_{L/R}^\dagger (\vec{r}) \Psi_{L/R}(\vec{r}),
\end{equation}

with $V_\perp(\vec{r})$ the cavity photonic pseudo potential and $\Psi_{L/R}(\vec{r})$ the fermion field operator on dot $L/R$. The parametric drive $V_g \cos(\drivepuls t + \phiRadiation)$, with $V_g$ the voltage amplitude applied to the gate, modulates the energies in both dots and we treat it semi-classically, giving

\begin{equation}
    \hat{H}_{mod} = q V_g \left[ \cos(\drivepuls t + \phiRadiation) \left( \alpha_{L} \ketbra{L}  + \alpha_{R} \ketbra{R} \right) + \cos(\drivepuls t + \phiTunnel) \left( \beta_t \ketbra{L}[R] + \beta_t \ketbra{R}[L] \right) \right],
\end{equation}

with $\alpha_{L}=\alpha_{LL}$ and $\alpha_{R} = \alpha_{RL}$ the electrostatic lever-arms from the left gate to the left dot and from the left gate to the right dot respectively, in our particular implementation, and $\beta_t$ is an effective lever-arm to the tunnel coupling term. We then diagonalize $\hat{H}_{\rm dqd}$, giving eigenenergies $E_\pm =\frac{\epsilon_L + \epsilon_R}{2} \pm \sqrt{\detuning^2 + 4 \tunnel^2}$, where the detuning energy is $\detuning=\epsilon_L - \epsilon_R$ and corresponding eigenvectors

\begin{align*}
	\Ket{\psi_{-}} & = \cos\left(\frac{\theta}{2} \right) \Ket{L} + \sin\left(\frac{\theta}{2} \right) \Ket{R} \\
	\Ket{\psi_{+}} & = \sin\left(\frac{\theta}{2} \right) \Ket{L} - \cos\left(\frac{\theta}{2} \right) \Ket{R}
\end{align*}

where the pairing angle $\theta$ is defined as $\tan \left( \theta \right)=\frac{2 \tunnel}{\detuning}$. Note that we can re-express the detuning as a function of the plunger gates voltages $V_{PL}$ and $V_{PR}$ as $\detuning = q \left[(\alpha_{LL} V_{PL} +  \alpha_{LR}) V_{PR}) - (\alpha_{RL} V_{PL} +  \alpha_{RR}) V_{PR}) \right] = q (\beta_L V_{PL} - \beta_R V_{PR})$. This defines the electrostatic lever-arm $\beta_L = \alpha_{LL} - \alpha_{RL}$ used in the main text.

We then rewrite $\hat{H}$ in the $\{\Ket{\psi_{-}}, \Ket{\psi_{+}}\}$ basis where the modified terms are

\begin{subequations}
\begin{align}
    \hat{H}_{dqd} & = \frac{\hbar \qubitpuls}{2} \hat{\sigma}_z \\ 
    \hat{H}_{int} & = \hbar \left(\ann{a} + \cre{a} \right) \left[ \gbare \sin \theta (\hat{\sigma}_{+} + \hat{\sigma}_{-}) + g_\Sigma \hat{\sigma}_0 -\gbare \cos \theta \hat{\sigma}_z \right] \\
    \hat{H}_{mod} & =  q V_g \left\{ \left[\cos(\drivepuls t + \phiRadiation) \beta_L \sin \theta - \cos(\drivepuls t + \phiTunnel) \beta_t \cos \theta \right] (\hat{\sigma}_{+} + \hat{\sigma}_{-}) \right. \nonumber \\
    & \;\;\; + \left. \beta_\Sigma \hat{\sigma}_0 - \left[\cos(\drivepuls t + \phiRadiation) \beta_L \cos \theta + \cos(\drivepuls t + \phiTunnel) \beta_t \sin \theta\right] \hat{\sigma}_z \right\},
\end{align}
\end{subequations}

where we have defined $\gbare = \frac{g_L - g_R}{2}$ the bare dipole-photon coupling, $g_\Sigma = \frac{g_L + g_R}{2}$ and $\beta_\Sigma = \alpha_L + \alpha_R=\alpha_{LL} + \alpha_{RL}$. We now turn to the equations of motion (EOM) of the various operators. If the number of photons $n_{\rm ph}$ is large enough (typically $n_{\rm ph}=\langle \numop{a} \rangle \gtrsim 10$) as is the case in this work, we have $\ann{a} \approx \langle a \rangle = \bar{a}e^{-i \drivepuls{} t}$ and $\dot{\ann{a}}=-i \drivepuls{} \ann{a}$ where we have done first a semi-classical approximation and second the resonant approximation~\cite{Cottet2017}. With this, we have

\begin{subequations}
\begin{align}
	\partial_t \langle \ann{a} \rangle = & -i \cavpuls \avg{\ann{a}} - i |A_{in}| e^{-i (\drivepuls t + \phi_{in})} - \frac{\kappa}{2} \avg{\ann{a}} - i \left[ \gbare \sin \theta (\avg{\sigma_{+}} + \avg{\sigma_{-}}) + g_\Sigma - \gbare \cos \theta \avg{\sigma_z } \right]\label{eq:SM:EOM_a} \\ 
    \partial_t \avg{\sigma_{-}} = & - [\Gamma_2 + i \qubitpuls] \avg{\sigma_{-}} + i \gbare \sin \theta \avg{(a + a^\dagger) \sigma_{z}} - 2 i \gbare \cos \theta \avg{(a + a^\dagger) \sigma_{-}} \nonumber  \\
	& + i \frac{q V_g}{\hbar} \left\{\left[\cos(\drivepuls t + \phiRadiation) \beta_L \sin \theta - \cos(\drivepuls t + \phiTunnel) \beta_t \cos \theta \right] \avg{\sigma_{z}} \right. \nonumber \\
    & \left. - 2 i \left[\cos(\drivepuls t + \phiRadiation) \beta_L \cos \theta + \cos(\drivepuls t + \phiTunnel) \beta_t \sin \theta\right] \avg{\sigma_{-}} \right\}\label{eq:SM:EOM_sigma_minus_mod} \\
	\partial_t \avg{\sigma_{+}} = & - [\Gamma_2 - i \qubitpuls] \avg{\sigma_{+}} - i \gbare \sin \theta \avg{(a + a^\dagger) \sigma_{z}} + 2 i \gbare \cos \theta \avg{(a + a^\dagger) \sigma_{+}} \nonumber \\
	& - i \frac{q V_g}{\hbar} \left\{ \left[\cos(\drivepuls t + \phiRadiation) \beta_L \sin \theta - \cos(\drivepuls t + \phiTunnel) \beta_t \cos \theta \right] \avg{\sigma_{z}} \right. \nonumber \\
    & \left. + 2 i \left[\cos(\drivepuls t + \phiRadiation) \beta_L \cos \theta + \cos(\drivepuls t + \phiTunnel) \beta_t \sin \theta\right] \avg{\sigma_{-}} \right\} \label{eq:SM:EOM_sigma_plus_mod},
\end{align}
\end{subequations}

where we have introduced the decoherence rate $\Gamma_2$ of the dipole and the cavity photon loss rate $\kappa$ following the standard procedure~\cite{blais2021}. In our case, $\hat{\sigma}_z$ is quasi-static and the number of photons is low enough that we do not need to account for out-of-equilibrium drive of $\avg{\sigma_z}$. In the steady-state, we thus get 

\begin{subequations}\label{eq:SM:sigma_SS}
\begin{align}
	\avg{\sigma_{-}} = & \left[ \frac{\gtrans \avg{a} + \frac{q V_g}{2 \hbar} e^{-i \drivepuls t} \left[e^{-i \phiRadiation} \beta_L \sin \theta - e^{-i \phiTunnel} \beta_t \cos \theta \right] }{\detqd - i \Gamma_2} \right. \\ \nonumber 
    & + \left. \frac{\gtrans \avg{a^\dagger} + \frac{q V_g}{2 \hbar} e^{ i \drivepuls t}\left[e^{i \phiRadiation} \beta_L \sin \theta - e^{i\phiTunnel} \beta_t \cos \theta \right] }{\Sigma_{qd} - i \Gamma_2} \right] \avg{\sigma_{z}} \\ 
	\avg{\sigma_{+}} = & \left[ \frac{\gtrans \avg{a} + \frac{q V_g}{2 \hbar} e^{-i \drivepuls t} \left[e^{-i \phiRadiation} \beta_L \sin \theta - e^{-i \phiTunnel} \beta_t \cos \theta \right] }{\Sigma_{qd} + i \Gamma_2} \right. \\ \nonumber
    & + \left. \frac{\gtrans \avg{a^\dagger} + \frac{q V_g}{2 \hbar} e^{ i \drivepuls t} \left[e^{i  \phiRadiation} \beta_L \sin \theta - e^{i \phiTunnel} \beta_t \cos \theta \right] }{\detqd + i \Gamma_2} \right] \avg{\sigma_{z}},
\end{align}
\end{subequations}

with $\detqd=\qubitpuls - \drivepuls$ and $\Sigma_{qd} = \qubitpuls + \drivepuls$. We keep the Bloch-Siegert terms in $\Sigma_{qd}$ because in our experimental realization, we have $\detqd \gtrsim \drivepuls$, therefore the Bloch-Siegert terms have a non-negligible contribution as $\Sigma_{qd}$ is of the same order of magnitude as $\detqd$. We add the modulation of $\gbare=\bar{g}_0 + \tilde{g}_0 \cos (\drivepuls t + \phiLong)$ as in ref.~\cite{Didier2015}. We also add the parasitic drive of the cavity that is due to a capacitive leakage from the modulation gate to the cavity

\begin{equation}
    \hat{H}_{\rm drive} \rightarrow \hbar |\driveAmp| \left[e^{i (\drivepuls t + \phi_{in})} \ann{a} + e^{-i (\drivepuls t + \phi_{in})} \cre{a} \right] + \hbar |\driveFast| \left[e^{i (\drivepuls t + \phi_{in} + \phiInterf)} \ann{a} + e^{-i (\drivepuls t + \phi_{in} + \phiInterf)} \cre{a} \right],
\end{equation}

where we have defined the phase difference between the two drives as $\phiInterf$. Now, inserting eqs.~\eqref{eq:SM:sigma_SS} into eq.~\eqref{eq:SM:EOM_a} and keeping only terms that are resonant with $\hat{a}$, we find the steady state of the intra-cavity field

\begin{equation}\label{eq:SM:field_full}
	\bar{a} = \frac{\driveCav + A_{\parallel} + A_{mod}}{-\Delta_{cd} + i \frac{\kappa}{2}-\chi \avg{\sigma_z}},
\end{equation}

with

\begin{subequations}\label{eq:SM:numerator}
\begin{align}
    \chi & = \gtrans^2 \left( \frac{1}{\Delta_{\rm qd} - i \Gamma_2} + \frac{1}{\Sigma_{\rm qd} + i \Gamma_2} \right) \label{eq:SM:chi} \\
    \driveCav & = (|\driveAmp| + |\driveFast|e^{-i \delta \phi})e^{-i \phi_{in}}=|\driveAmp|e^{-i \phi_{in}}(1 + r e^{-i \delta \phi}) \label{eq:SM:driveAmp} \\
    A_{\parallel} & = -\tilde{g}_0 \cos \theta \avg{\sigma_z} = - \tilde{g}_\parallel \avg{\sigma_z} e^{-i \phiLong} \label{eq:SM:longitudinalAmp} \\
    A_{mod} & = \frac{q V_g}{2 \hbar} \frac{\chi}{\gbare} \left( \beta_L e^{-i \phiRadiation} - \beta_t \mathrm{cotan} \theta e^{-i \phiTunnel} \right) \avg{\sigma_z} = \frac{q V_g}{2 \hbar} \frac{\chi}{\gtrans} \left( \beta_L \sin \theta e^{-i \phiRadiation} - \beta_t \cos \theta e^{-i \phiTunnel} \right) \avg{\sigma_z} \label{eq:SM:modAmp}
\end{align}
\end{subequations}

where $\gtrans = \gbare \sin \theta$, $\tilde{g}_\parallel = \tilde{g}_0 \cos \theta$ and $r=\driveFast / \driveAmp$. Using the standard input-output formalism, we have $|\driveAmp|=\sqrt{\kappa_{in} P_{in} / \hbar \drivepuls}$ and $|\driveFast|=\sqrt{\kappa_{g} P_{g} / \hbar \drivepuls}$ with $\kappa_{in/g}$ the photon loss rate at the input/parasitic port and $P_{in/g}$ the microwave power at the input/parasitic port. We further have $V_g = \sqrt{Z_{g} P_{g}}$ with $Z_g$ the characteristic impedance of the line.

\clearpage

\section{Characterization of the cavity}

During all this work, we read the cavity during its ringdown. The reason is to collect the microwave signal from the cavity only and avoid any parasitic interfering microwave signal that does not enter the cavity (e.g. free space propagation in the sample holder, propagation in the PCB, etc...). These additional modes would otherwise impede the investigation of field interferences occurring inside the cavity. We characterize the cavity under this condition, driving it through the input port characterized by its photon loss rate $\kappa_{in}$ and through the parasitic port (the gates of the DQD) characterized by $\kappa_g$ as shown in Supplementary Figure~\ref{figSM:cavity} where we show the normalized transmitted amplitude $A/A_0$ in panel \figpanel{a} and the phase of the transmitted signal in panel \figpanel{b}. The value of $A_0$ is taken at resonance. As both transmission amplitudes and phase perfectly overlap when driving through both ports, we can conclude that we are indeed driving the same cavity mode through both ports. We find the resonance frequency of the cavity $\cavpuls/2 \pi=\SI{7.029456}{\giga\hertz}$ and the total loss rate $\kappa / 2 \pi = \SI{1.339}{\mega\hertz}$.

\begin{figure}[hbt]
\centering
\includegraphics[width=0.6\linewidth, angle=0]{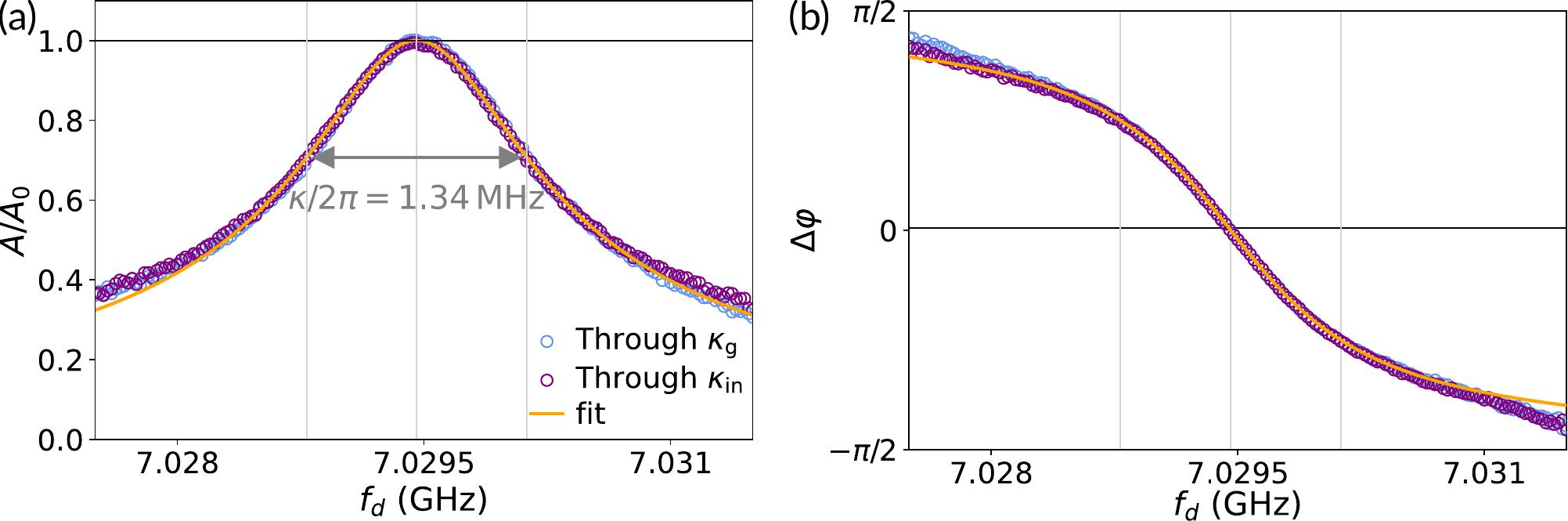}
\caption{Characterization of the cavity.
Normalized transmitted amplitude $A/A_0$ \figpanel{a} and phase of the transmitted signal\figpanel{b} through the cavity when driving from the input port of the cavity (purple circles) and from the parasitic port via the DQD gates (blue circles). The orange line is a fit.
}
\label{figSM:cavity}
\end{figure}

\clearpage

\section{Characterization of the first interdot transition parameters}

We calibrate the parameters of the interdot transition dipole by first performing a temperature dependence of the microwave response to extract the tunnel coupling $\tunnel$~\cite{Yu2023}, as shown in Supplementary Figure~\ref{figSM:temperature_2tone}\figpanel{a,b}. At equilibrium we have $\avg{\sigma_z}=-\tanh (\hbar \qubitpuls / 2 k_B T_e)$ with $k_B$ the Boltzmann constant and $T_e$ the electronic temperature. Fixing $\detuning=0$ at the center of the interdot transition, we get $\avg{\sigma_z}=-\tanh (\tunnel / k_B T_e)$. In the regime of the studied interdot transition, we have $\chi \avg{\sigma_z} \ll \kappa / 2$ so that $\Delta \readoutPhase \approx \frac{2}{\kappa}\mathrm{Re}[\chi]\avg{\sigma_z} \approx -K \times \tanh (\tunnel / k_B T_e)$ where $K=\frac{2}{\kappa}\mathrm{Re}[\chi]$ is constant with temperature, and taken at $\detuning=0$. The recorded values of $\Delta \readoutPhase(\detuning=0)$ as a function of the temperature $T$ of the mixing chamber (MC) are shown in the Supplementary Figure~\ref{figSM:temperature_2tone}\figpanel{c} along the corresponding fit. We considered a base electronic temperature of $T_0 = \SI{50}{\milli\kelvin}$ at $T=\SI{20}{\milli\kelvin}$, giving an electronic temperature $T_e=\sqrt{T_0^2 + T^2}$. We obtain $\tunnel\approx\SI{8.67}{\giga\hertz}$. We further confirm this value by performing a two-tone spectroscopy and find a resonance at \SI{17.56}{\giga\hertz} which disperses as $\sqrt{\detuning^2 + 4 \tunnel^2}$ as shown in Supplementary Figure~\ref{figSM:temperature_2tone}\figpanel{d}, corresponding to $\tunnel=\SI{8.78}{\giga\hertz}$. We keep this value for $\tunnel$. Note that the bright spot at around \SI{17.6}{\giga\hertz} does not show any dispersion and is probably due to a parasitic mode in the device.

Once the tunnel coupling is determined, we determine the other relevant parameters $\gbare$, $\beta_L$, $\beta_R$ and $\Gamma_2$ by fitting simultaneously the $\Delta A / A$ and $\Delta \readoutPhase$ quantities as a function of $\detuning$ for a given MC temperature $T$. For this we fix $\tunnel$, which sets the energy scale. We also fix the ratio $\beta_L / \beta_R=0.8$ which is inferred from the interdot transition main axis (along the interdot transition, the condition $\mu_L=\mu_R$ with $\mu_\alpha$ the chemical potential of dot $\alpha$ is satisfied) as can measured from Fig.~2\figpanel{b} of the main text. We convert the plunger gater voltages to detuning energy as $\detuning=q(\beta_L V_{PL} - \beta_R V_{PR})$, taking $q=-|e|$. In the fit, we thus have $\tunnel$ fixed and the constraint $\beta_L / \beta_R=0.8$. The remaining fitting parameters are $\beta_L$, $\gbare$ and $\Gamma_2$.

\begin{figure*}[hbt]
\centering
\includegraphics[width=0.6\linewidth, angle=0]{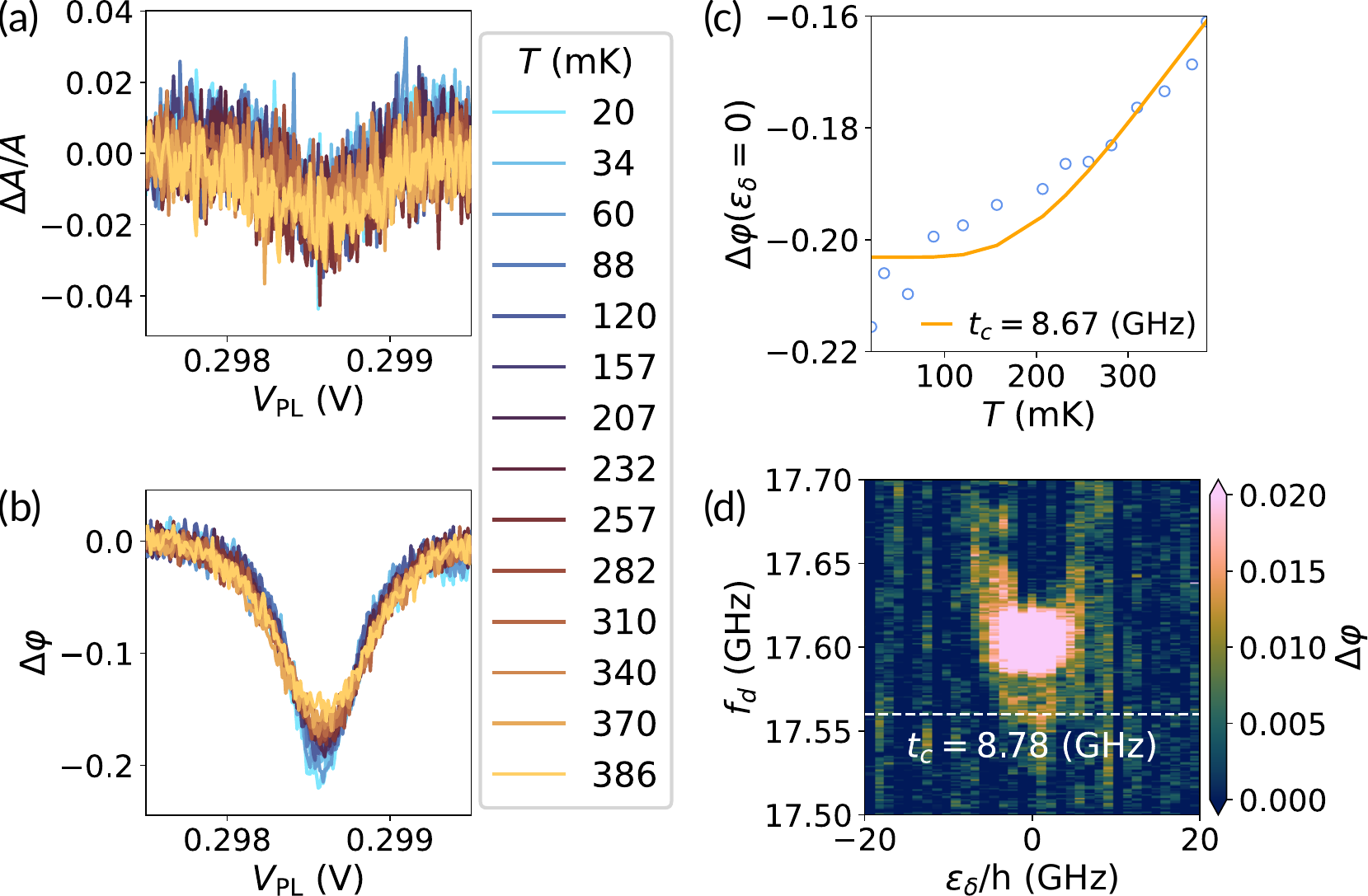}
\caption{Characterization of the first interdot transition discussed in the main text. 
Relative amplitude variation $\Delta A / A$ \figpanel{a} and phase variation $\Delta \readoutPhase$ \figpanel{b} as a function of the plunger gate voltage $V_{PL}$ along the detuning axis indicated by a black arrow in Fig.~2\figpanel{b} of the main text and for different MC temperatures. \figpanel{c} $\Delta \readoutPhase(\detuning=0)$ as a function of MC temperature $T$ (blue circles) and fit to the formula discussed in the text in orange. \figpanel{d} Two-tone spectroscopy showing $\Delta \readoutPhase$ as a function of detuning $\detuning$ and second tone drive frequency $f_d$.
}
\label{figSM:temperature_2tone}
\end{figure*}

\clearpage
 
\section{Antisymmetric part of the microwave field under parametric modulation of the detuning}

The odd part in $\detuning$ of the cavity field signal presented in Fig.~3 of the main text allows us to determine the contributions of the $\beta_t$ term in $A_{mod}$ (eq.~\eqref{eq:SM:modAmp}) and the longitudinal coupling term $A_\parallel$ (eq.~\eqref{eq:SM:longitudinalAmp}). The odd part of the data and corresponding simulations are shown in Supplementary Figure~\ref{figSM:stick1_odd}. We observe a very good qualitative agreement when setting $\beta_t = 0.03 \beta_L$, $\phi_t = \phi_m$ and $g_\parallel=0$ as discussed in the main text. The quantitative agreement is less good than for the raw data (mainly the even part) and can be attributed to the small signal amplitude resulting in a poor SNR. In addition, the extraction of the odd part in the data is strongly affected by the symmetry axis set by $\detuning=0$ and any small deviation from this point can lead to a change of amplitude of the odd part contribution which is not negligible compared to the signal amplitude. Part of the quantitative discrepancy observed between the data and the theory can therefore be explained by the resolution in $\detuning$.

\begin{figure*}[hbt]
\centering
\includegraphics[width=0.5\linewidth, angle=0]{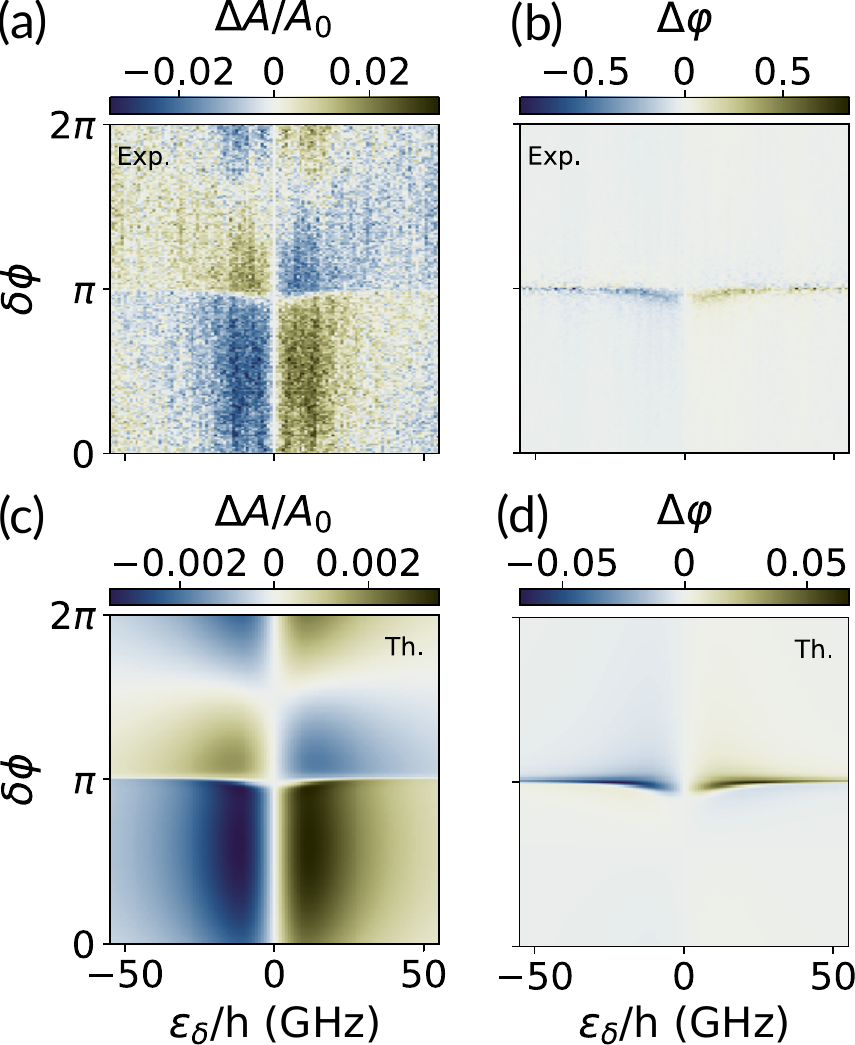}
\caption{Antisymmetric part of the modulated signal.
Odd part in $\detuning$ of the relative amplitude variation $\Delta A / A_0$ \figpanel{a} and phase variation $\Delta \readoutPhase$ \figpanel{b}, extracted from the data presented in Fig.~3 of the main text. The corresponding simulations based on eq.~\eqref{eq:SM:field_full} are respectively shown in panels~\figpanel{c,d}.
}
\label{figSM:stick1_odd}
\end{figure*}

\clearpage

\section{Modulated signal at different ratio of direct drives amplitudes}

\begin{figure*}[hbt]
\centering
\includegraphics[width=0.4\linewidth, angle=0]{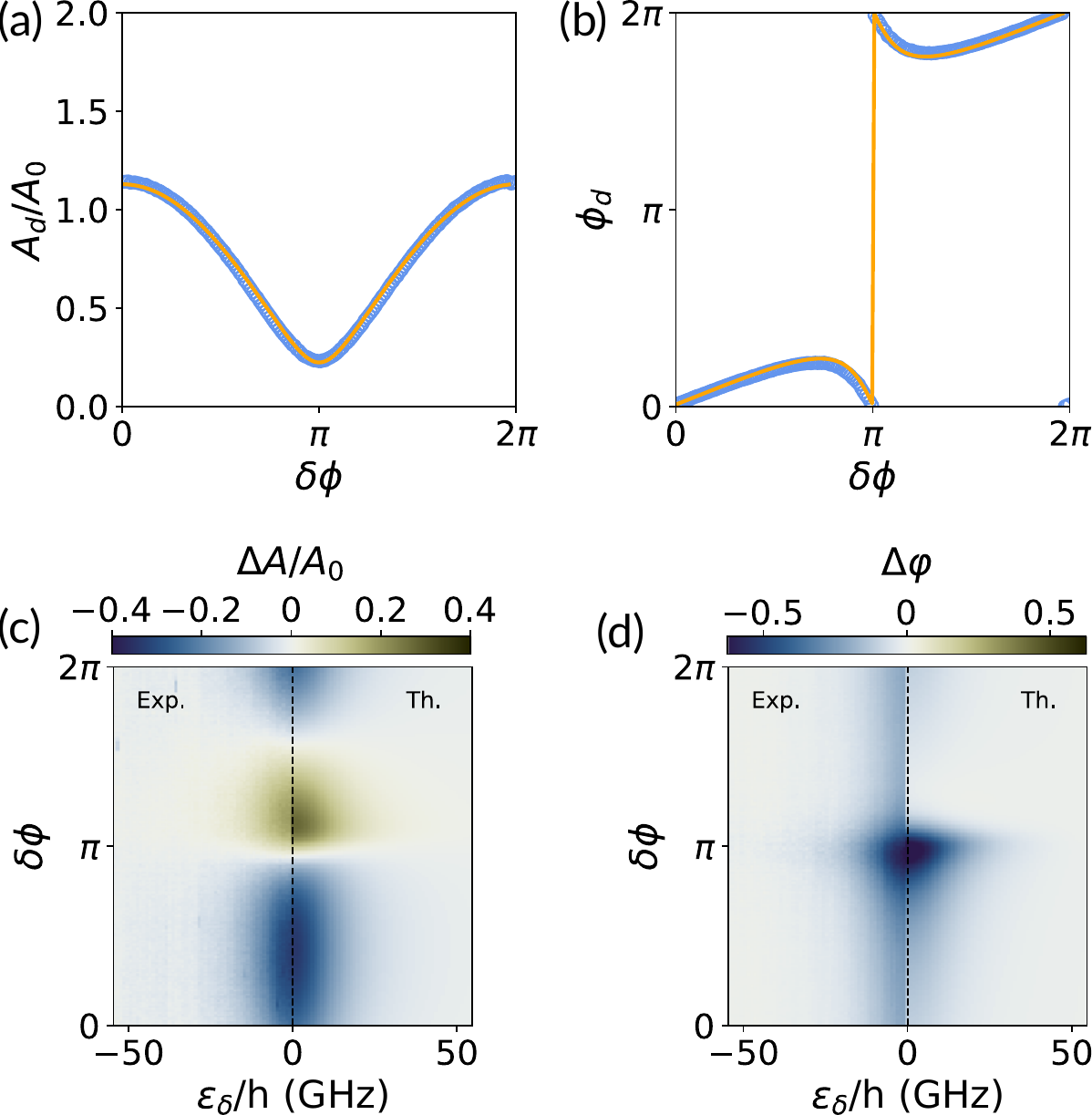}
\caption{Dynamical readout of the interdot transition for $r=0.74$.
Normalized output cavity field amplitude $A_d/A_0$ \figpanel{a} and phase $\phi_d$ of $A_d$ \figpanel{b}, for $\chi=0$ ($\detuning \gg \tunnel$) and $r\approx 0.74$, as a function of the phase difference $\phiInterf$ between the input ($A_{in})$ and parasitic ($A_g$) drives. \figpanel{c} Relative field amplitude variation of the interdot transition taken along the detuning axis (black arrow in Fig.~2\figpanel{b} of the main text) and varying the drives interference phase $\phiInterf$ at constant $r\approx 0.74$. Experimental data (left side) and theory (right side) are displayed on the same plot. \figpanel{d} Same as \figpanel{c} but for the quantity $\Delta \readoutPhase$.
}
\label{figSM:r0p6}
\end{figure*}

\begin{figure*}[hbt]
\centering
\includegraphics[width=0.4\linewidth, angle=0]{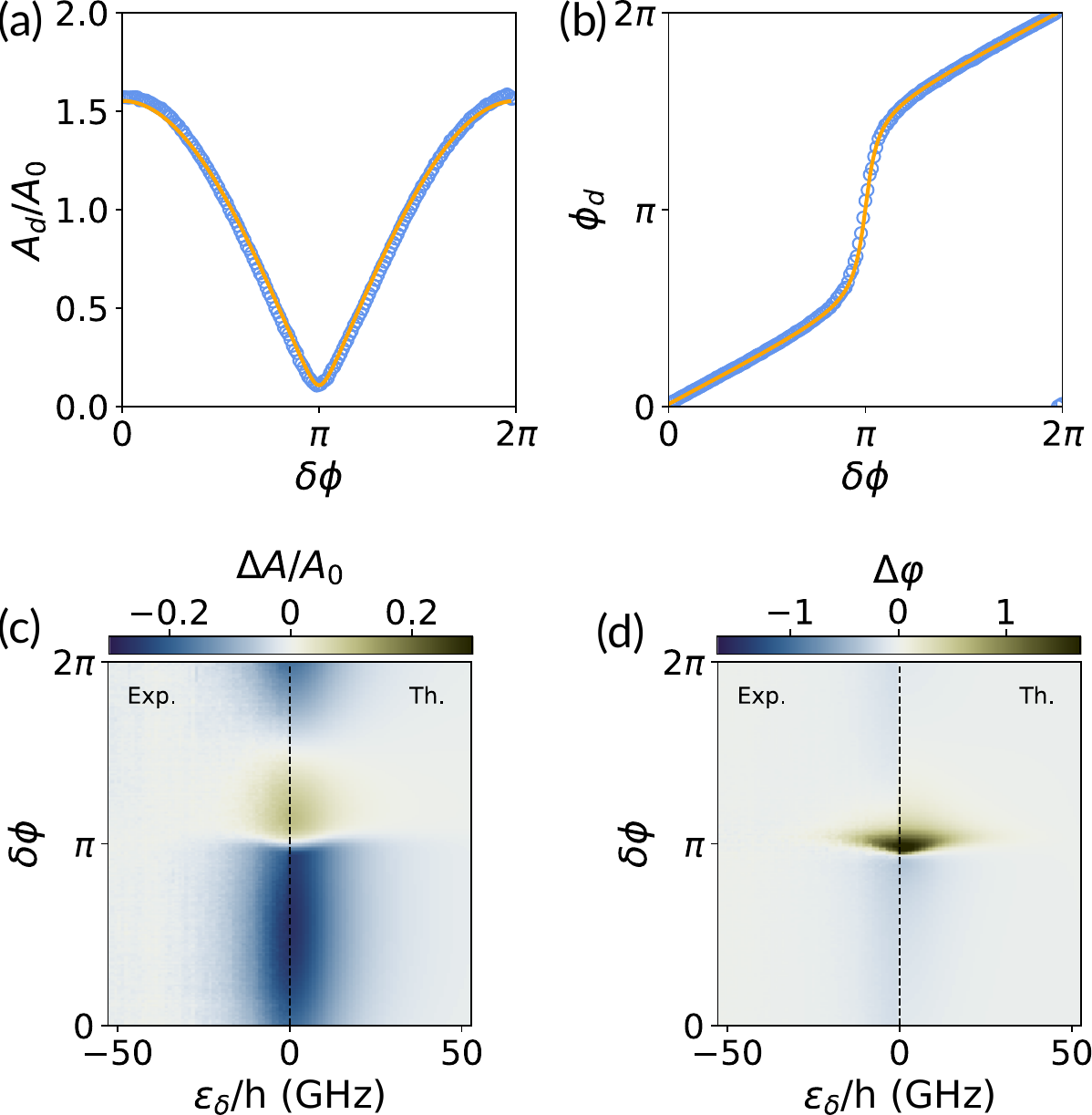}
\caption{Dynamical readout of the interdot transition for $r=1.21$.
Normalized output cavity field amplitude $A_d/A_0$ \figpanel{a} and phase $\phi_d$ of $A_d$ \figpanel{b}, for $\chi=0$ ($\detuning \gg \tunnel$) and $r\approx 1.21$, as a function of the phase difference $\phiInterf$ between the input ($A_{in})$ and parasitic ($A_g$) drives. \figpanel{c} Relative field amplitude variation of the interdot transition taken along the detuning axis (black arrow in Fig.~2\figpanel{b} of the main text) and varying the drives interference phase $\phiInterf$ at constant $r\approx 1.21$. Experimental data (left side) and theory (right side) are displayed on the same plot. \figpanel{d} Same as \figpanel{c} but for the quantity $\Delta \readoutPhase$.
}
\label{figSM:r1p2}
\end{figure*}

Here we present the data and simulations of the radiated field in the cavity as a function of $\detuning$ and $\phiInterf$ for the case when $r=|\driveFast| / |\driveAmp| = 0.74$ (Supplementary Figure~\ref{figSM:r0p6}) and $r=|\driveFast| / |\driveAmp| = 1.21$ (Supplementary Figure~\ref{figSM:r1p2}), showing again excellent quantitative agreement. This is for the first interdot transition discussed in the main text.

\clearpage

\section{Details of the ``second'' interdot transition}

Here we detail the characterization of the second interdot dipole transition discussed in the main text (Fig.~3\figpanel{i--h}). Although referred to as the ``second'' transition, this dataset was chronologically acquired after the ``third'' interdot transition discussed in the manuscript and characterized in the next supplementary section.

\begin{figure*}[hbt]
\centering
\includegraphics[width=0.6\linewidth, angle=0]{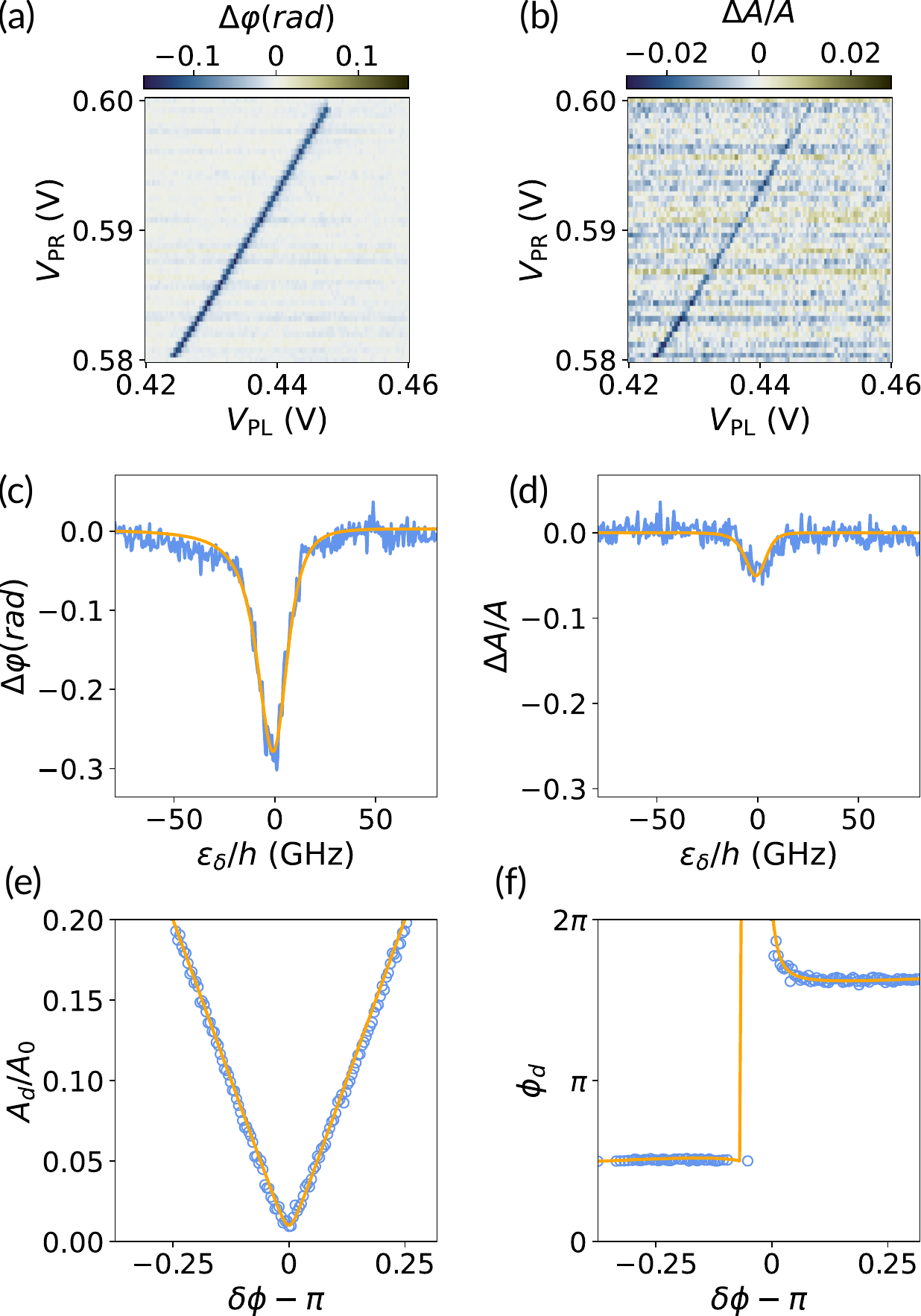}
\caption{Second interdot transition.
Phase variation $\Delta \readoutPhase$ \figpanel{a} and relative amplitude variation $\Delta A / A_{\rm off}$ \figpanel{b} of the cavity field probed at $f_c$ across the second interdot transition, as a function of the left and right plunger gate voltages. Cuts in detuning $\detuning$ across the intedot transition at lower readout drive power showing the phase contrast $\Delta \readoutPhase$ \figpanel{c} and relative amplitude variation $\Delta A / A$ \figpanel{d}. \figpanel{e} Normalized output cavity field amplitude $A_d/A_0$ and phase of the output field \figpanel{f}, for $\chi=0$ ($\detuning \gg \tunnel$) and $r\approx 1$, as a function of the phase difference $\phiInterf$ between the input ($A_{in})$ and parasitic ($A_g$) drives.
}
\label{figSM:stick3_charac}
\end{figure*}

Supplementary Figure~\ref{figSM:stick3_charac}(a,b) presents the standard dispersive readout characterization of the interdot transition. Panels~\figpanel{c,d} show cuts along the detuning axis ($\detuning$) across this transition. From the slope in the $V_{\rm PL}$--$V_{\rm PR}$ plane satisfying $\mu_L = \mu_R$, we extract the lever-arm ratio $\beta_L/\beta_R=0.83$, consistent with the ratio previously found for the first interdot transition. The reduced phase contrast $\Delta\varphi$ in panel~\figpanel{a} compared to panel~\figpanel{c} results from a larger (by a factor $\sim$6) readout power used for faster acquisition, leading to photon-induced excitation of the dipole transition and thus reducing $\avg{\sigma_z}$~\cite{Viennot2014}. Therefore, we use the low-power data (Supplementary Fig.~\ref{figSM:stick3_charac}(c,d)) to extract the dipole parameters: $\tunnel\approx\SI{6.3}{\giga\hertz}$, $\gbare\approx\SI{29}{\mega\hertz}$, and an upper bound $\Gamma_2\leq\SI{300}{\mega\hertz}$. For consistency, we set $\beta_L=-0.091$ (as measured for the first transition), justified by the similar gate-voltage parameter space and identical $\beta_L$ measured on the third transition.

\begin{figure*}[hbt]
\centering
\includegraphics[width=0.6\linewidth, angle=0]{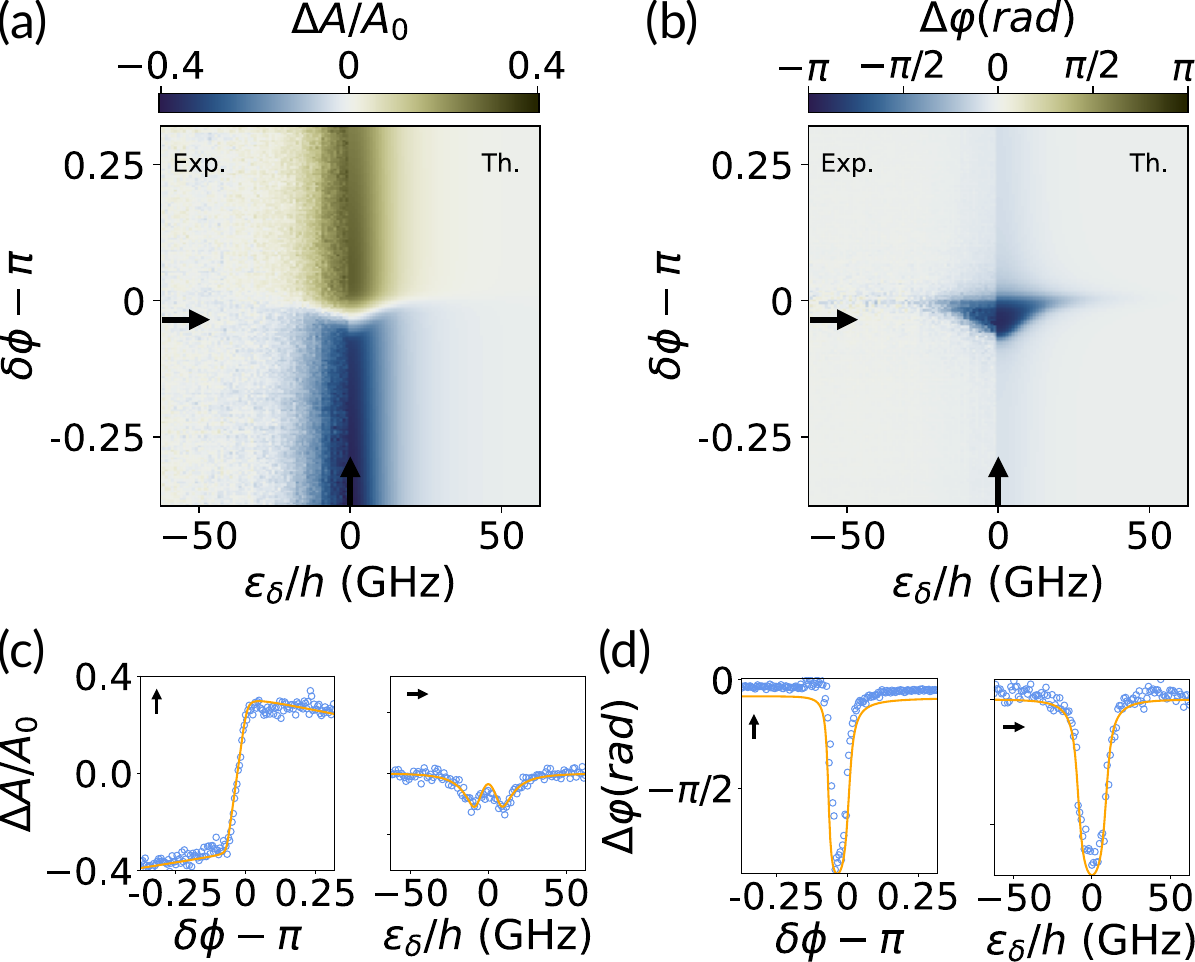}
\caption{Second interdot transition.
\figpanel{a} Relative field amplitude variation $\Delta A / A_0$ of the second interdot transition along the detuning axis and varying the drives interference phase $\phiInterf$ at constant $r = 0.988$. Experimental data (left side) and theory (right side) are displayed on the same plot. \figpanel{b} Same as \figpanel{a} but for the quantity $\Delta \readoutPhase$. Respective cuts along $\phiInterf$ at zero detuning are shown in \figpanel{c,e} and cuts along $\detuning$ at $\phiInterf\approx0.99 \pi$ are shown in \figpanel{d,f}. Data are shown by blue circles and simulations by orange lines.
}
\label{figSM:stick3_radiation}
\end{figure*}

We next implement the parametric drive readout. From data shown in Supplementary Fig.~\ref{figSM:stick3_charac}\figpanel{e,f}, we determine the amplitude ratio between the input and parasitic drives as $r\approx 0.988$. The parametric readout results (Supplementary Fig.\ref{figSM:stick3_radiation}), analogous to Fig.~3 in the main text, show the relative amplitude and phase variations as functions of detuning and drive interference phase $\phiInterf$. The measurements display again very good agreement with theoretical simulations, from which we extract parameters $\phi_m \approx 0.49 \times \pi$ and $\beta_t \approx 0.003$.

\clearpage

\section{Details of the ``third'' interdot transition}

We present here the details of the third interdot dipole transition discussed in the main text and measured for the optimization of the SNR presented in Fig.~4 of the main text.

First we show in Supplementary Figure~\ref{figSM:stick2_charac} the measurement of the dipole transition with conventional dispersive readout in panels~\figpanel{a,b} and the intra-cavity field control close to $r=1$ in panels~\figpanel{c,d}. The phase contrast $\Delta \readoutPhase$ across the transition is small with a variation of $\approx \SI{-0.05}{\radian}$ and no amplitude variation. This indicates a transition very deep in the dispersive regime. We also note from Supplementary Figure~\ref{figSM:stick2_charac}\figpanel{a} that the condition $\mu_L = \mu_R$ is fulfilled for $\beta_L/\beta_R=1$, which is then used as a constraint for the subsequent fits.

\begin{figure*}[hbt]
\centering
\includegraphics[width=0.6\linewidth, angle=0]{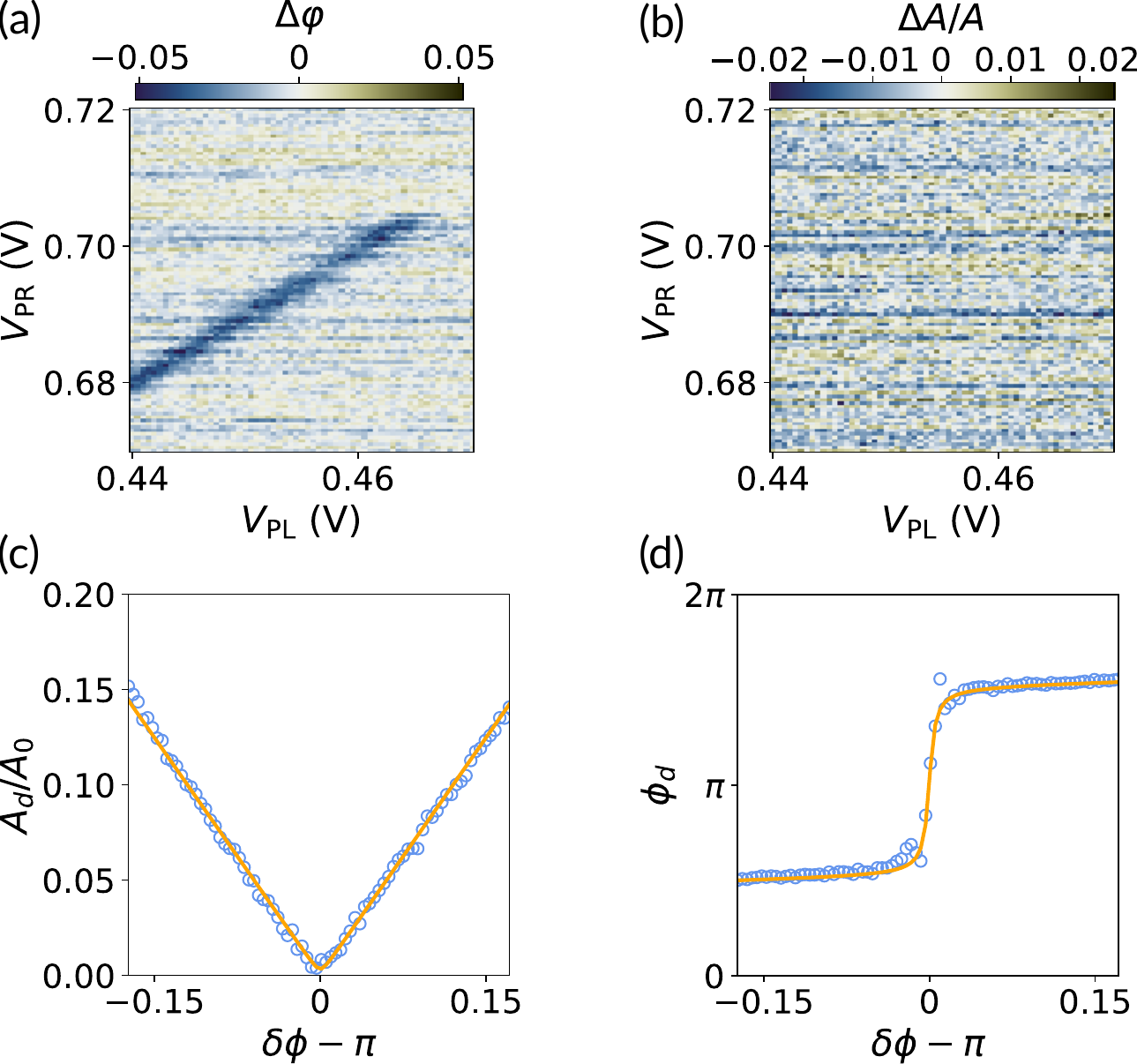}
\caption{Second interdot transition.
Phase variation $\Delta \readoutPhase$ \figpanel{a} and relative amplitude variation $\Delta A / A_{\rm off}$ \figpanel{b} of the cavity field probed at $f_c$ across the second interdot transition, as a function of the left and right plunger gate voltages. \figpanel{c} Normalized output cavity field amplitude $A_d/A_0$ and phase of the output field \figpanel{d}, for $\chi=0$ ($\detuning \gg \tunnel$) and $r\approx 1$, as a function of the phase difference $\phiInterf$ between the input ($A_{in})$ and parasitic ($A_g$) drives.
}
\label{figSM:stick2_charac}
\end{figure*}

We present in Supplementary Figure~\ref{figSM:stick2_radiation} the parametric readout of the transition when modulating $\detuning$ at the cavity frequency, as a function of $\detuning$ and the phase between the two drive tones $\phiInterf$ similarly to Fig.~3 of the main text. We did not characterize the dipole transition by performing a temperature dependence but instead obtained the parameters of this interdot dipole transition by performing a joint fit on the data $\Delta \readoutPhase$ and $\Delta A / A_0$ of Fig.~4\figpanel{a,b} of the main text and of Supplementary Figure~\ref{figSM:stick2_radiation}\figpanel{a,b}. We fix $r=1.004$ (from the Supplementary Figure~\ref{figSM:stick2_charac}\figpanel{c,d}), $\sqrt{\kappa_{g}}\times V_{g}$ (from the fit of the first dipole transition) and $\beta_L/\beta_R=1$. The fit is highly constrained by the two different data maps. We find $\beta_L=-0.087$, very close to the the value of the first dipole transition, $\gbare \approx \SI{22}{\mega\hertz}$, $\tunnel\approx \SI{20.8}{\giga\hertz}$, $\phi_m \approx 0.49 \times \pi$ and $\beta_t \approx 0.01 \beta_L$, also close to the value found for the first dipole transition.

\begin{figure*}[hbt]
\centering
\includegraphics[width=0.6\linewidth, angle=0]{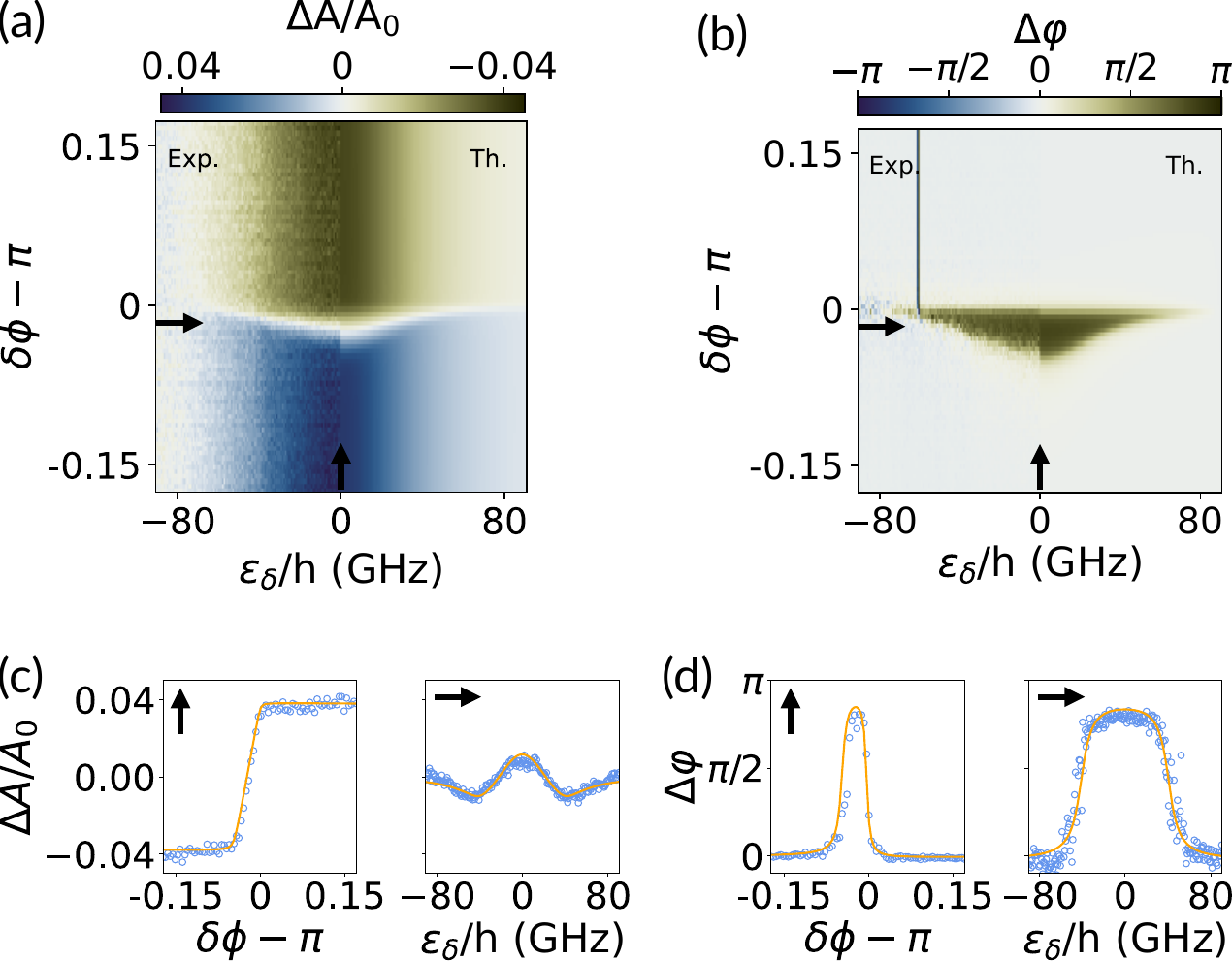}
\caption{Second interdot transition.
\figpanel{a} Relative field amplitude variation $\Delta A / A_0$ of the second interdot along the detuning axis and varying the drives interference phase $\phiInterf$ at constant $r = 1.004$. Experimental data (left side) and theory (right side) are displayed on the same plot. \figpanel{b} Same as \figpanel{a} but for the quantity $\Delta \readoutPhase$. Respective cuts along $\phiInterf$ at zero detuning are shown in \figpanel{c,e} and cuts along $\detuning$ at $\phiInterf\approx0.99 \pi$ are shown in \figpanel{d,f}. Data are shown by blue circles and simulations by orange lines.
}
\label{figSM:stick2_radiation}
\end{figure*}

\makeatother

\end{document}